\documentclass[12pt]{article}
\usepackage{amsmath} 
\usepackage{graphicx} 
\newcommand\as{\alpha_{\mathrm{S}}} 
\newcommand\f[2]{\frac{#1}{#2}} 
\newcommand{\plusd}[3][+]{{\left(\frac{#2}{#3}\right)_{#1}}}
\newcommand{\T}{\bom T}
\def\bom#1{{\mbox{\boldmath $#1$}}} 
\def\to{\rightarrow}
 
\def\nn{\nonumber}

\def\ep{\epsilon}
\def\ms{${\overline {\rm MS}}$}
\def\xw{x_\omega}

\def\beq{\begin{equation}}
\def\eeq{\end{equation}}
\def\beeq{\begin{eqnarray}}
\def\eeeq{\end{eqnarray}}

\def\bP{{\bf P}}
\def\bp{{\bf p}}

\begin{document} 
\begin{titlepage}
\renewcommand{\thefootnote}{\fnsymbol{footnote}}
\begin{flushright}
ZU-TH 09/13
\end{flushright}
\vspace*{2cm}

\begin{center}
{\Large \bf Soft-gluon resummation for single-particle inclusive\\[0.2cm] hadroproduction at high transverse momentum}
\end{center}

\par \vspace{2mm}
\begin{center}
{\bf Stefano Catani$^{(a)}$,
Massimiliano Grazzini$^{(b)}$\footnote{On leave of absence from INFN, Sezione di Firenze, Sesto Fiorentino, Florence, Italy.}}
and
{\bf Alessandro Torre$^{(b)}$}

\vspace{5mm}

$^{(a)}$ INFN, Sezione di Firenze and Dipartimento di Fisica e Astronomia,\\ 
Universit\`a di Firenze,
I-50019 Sesto Fiorentino, Florence, Italy\\

$^{(b)}$ Institut f\"ur Theoretische Physik, Universit\"at Z\"urich, CH-8057 Z\"urich, Switzerland

\vspace{5mm}

\end{center}

\par \vspace{2mm}
\begin{center} {\large \bf Abstract} \end{center}
\begin{quote}
\pretolerance 10000

We consider the cross section for one-particle 
inclusive production at high transverse 
momentum in hadronic collisions. 
We present the all-order 
resummation formula that controls the logarithmically-enhanced 
perturbative QCD contributions to the partonic cross section in the
threshold region, at fixed rapidity of the
observed parton (hadron).
The explicit resummation up to next-to-leading logarithmic accuracy is
supplemented with the computation of
the general structure of the near-threshold
contributions to the next-to-leading order cross section.
This next-to-leading order computation allows us to extract 
the one-loop hard-virtual amplitude
that enters into the resummation formula.
This is a necessary ingredient to explicitly extend the soft-gluon resummation
beyond the next-to-leading logarithmic accuracy.
These results equally apply to both spin-unpolarized and spin-polarized 
scattering processes.

\end{quote}

\vspace*{\fill}
\begin{flushleft}
May 2013

\end{flushleft}
\end{titlepage}

\setcounter{footnote}{1}
\renewcommand{\thefootnote}{\fnsymbol{footnote}}

\section{Introduction}
\label{sec:intro}

A well known feature of QCD is that perturbative computations for
hard-scattering processes are sensitive to soft-gluon effects.
These effects manifest themselves when the considered observable is computed
close to its corresponding boundary of the phase space. In these kinematical
regions, real radiation is strongly inhibited
and the cancellation of infrared singular terms between  virtual and real
emission  contributions is unbalanced.
This leads to large logarithmic terms that can invalidate the (quantitative)
reliability of the order-by-order perturbative expansion in powers of the QCD
coupling $\as$.
These large logarithmic terms have to be evaluated at sufficiently-high
perturbative orders and,
whenever it is possible, they should be resummed to all orders in 
QCD perturbation theory.

In the context of hadron--hadron collisions, 
a class of soft-gluon sensitive observables
is represented by inclusive hard-scattering cross sections in kinematical
configurations that are close to (partonic) threshold.
Typical examples are the cross sections for the production 
of Drell-Yan lepton pairs and Higgs bosons.
In these cases, where only two QCD partons
enter the hard-scattering subprocess at the Born level,
the soft-gluon resummation formalism
was established long ago \cite{Sterman:1986aj, Catani:1989ne, Catani:1990rp},
and explicit resummed results have been obtained up to 
next-to-next-to-leading logarithmic (NNLL) accuracy 
\cite{Vogt:2000ci, Catani:2001ic, Catani:2003zt}, and including 
still higher-order logarithmic terms that have been explicitly computed
\cite{Moch:2005ky, Laenen:2005uz}.
The case of cross sections that are produced by Born-level hard scattering
of three and four (or more) coloured partons is very important from the 
phenomenological viewpoint, 
and it is much more complex on the theoretical side.
Soft-gluon dynamics leads to non-trivial colour correlations and colour
coherence effects
that depend on the colour flow of the underlying partonic subprocess.
The general soft-gluon resummation formalism for inclusive cross sections in
these complex multiparton processes was developed in a series of papers
\cite{Kidonakis:1996aq, Bonciani:1998vc, Kidonakis:1998nf, Laenen:1998qw,
Catani:1998tm, Bonciani:2003nt}.
In recent years, techniques and methods 
of Soft Collinear Effective Theory (SCET) 
have also been developed and applied to resummation for inclusive 
cross sections near (partonic) threshold
\cite{Manohar:2003vb, Idilbi:2006dg, Becher:2006mr, Becher:2007ty, Ahrens:2008nc,
Beneke:2009rj, Becher:2009th}.

Some examples of relevant processes with three or four partons at the Born 
level are the direct production of prompt photons 
\cite{Laenen:1998qw, Catani:1998tm, Sterman:2000pt, Bolzoni:2005xn,
Becher:2009th}, 
vector boson \cite{Kidonakis:1999ur, Becher:2011fc} 
and Higgs boson \cite{de Florian:2005rr} production
at high transverse momentum,
production of heavy quarks 
\cite{Kidonakis:1996aq, Bonciani:1998vc, Laenen:1998qw, Almeida:2008ug, 
Beneke:2009rj, Czakon:2009zw, Ahrens:2010zv, Ahrens:2011mw}
and coloured supersymmetric particles (Ref.~\cite{Beenakker:2013mva}
and references therein) at hadron colliders, 
single top-quark production \cite{Kidonakis:2010tc, Zhu:2010mr},
jet \cite{Catani:1996yz, Kidonakis:1998bk, deFlorian:2007fv}
 and dihadron \cite{Almeida:2009jt, Kelley:2010fn} production,
and  single-hadron inclusive production 
in hadronic collisions \cite{deFlorian:2005yj}.
Soft-gluon resummation for single-hadron inclusive production in collisions 
of spin-polarized hadrons has been considered in Ref.~\cite{de Florian:2007ty}.

In this paper we 
consider the single-hadron inclusive cross section.
At sufficiently-large values of the hadron transverse momentum, 
the cross section for this process
factorizes into the convolution of the parton distribution
functions of the colliding hadrons with the (short-distance) partonic
cross section and with the fragmentation function of the triggered parton 
into the observed hadron. Since the single inclusive cross section can be 
easily measured by experiments in hadron collisions,
the process offers a relevant test of the QCD factorization picture. 
Conversely, measurements of the corresponding cross section as function of the
transverse momentum and at different collision energies
permit to extract quantitative information about the parton fragmentation 
(especially, the gluon fragmentation) function into the observed hadron, 
thus complementing the information obtained from hadron production in 
$e^+e^-$ and lepton-hadron collisions.

The next-to-leading order (NLO) QCD calculation of the cross section
for single-hadron inclusive production was completed long ago 
\cite{Ellis:1979sj, Aversa:1988fv, Aversa:1988vb}. 
Soft-gluon resummation of the logarithmically-enhanced contributions to the
partonic cross section was performed in Ref.~\cite{deFlorian:2005yj}.
The study of Ref.~\cite{deFlorian:2005yj} considers resummation for 
the transverse-momentum dependence of the cross section integrated over the
rapidity of the observed final-state hadron, and it explicitly resums
the leading-logarithmic (LL) and next-to-leading logarithmic (NLL) terms.
The results of the phenomenological studies (which combine NLL
resummation with the complete NLO calculation) in Ref.~\cite{deFlorian:2005yj}
indicate that the quantitative effect of resummation is 
rather large, especially in the kinematical configurations that are 
encountered in experiments at the typical energies of fixed-target collisions.

The content of the present paper aims at a twofold theoretical improvement on
resummation for single-hadron inclusive production:
we study soft-gluon resummation for the transverse-momentum cross section 
at fixed rapidity of the observed hadron (parton), and we extend the logarithmic
accuracy of resummation by
explicitly computing
a class of logarithmic terms beyond the NLL accuracy.
We first consider the structure of the NLO QCD corrections close to 
the partonic threshold.
In this kinematical region, the initial-state partons 
have just enough energy to produce the triggered final-state parton 
(that eventually fragments into the observed hadron) and a small-mass
recoiling jet, which is formed by soft and collinear partons.
We perform the NLO calculation by using soft and collinear approximations,
and we present a general expression for the logarithmically-enhanced terms 
(including the constant term)
that correctly reproduces the known NLO result.
Our NLO expression is directly factorized in colour space, and it allows us
to explicitly disentangle colour-correlation and colour-interference effects 
that contribute to soft-gluon resummation at NLL and NNLL accuracy.
We then consider 
the logarithmically-enhanced terms beyond the NLO.
We use the formalism of Ref.~\cite{Bonciani:2003nt}, and
we present the soft-gluon resummation formula that controls the logarithmic
contributions to the rapidity distribution of the transverse-momentum
cross section. The resummation formula is valid to arbitrary logarithmic
accuracy, and it is explicitly worked out up to the NLL level.
Finally, using our general expression of the NLO cross section, we determine
the one-loop hard-virtual amplitude that enters into 
the colour-space factorization structure of the resummation formula.
The colour interference between this one-loop amplitude and the NLL terms
explicitly determines an entire class of resummed contributions at NNLL accuracy.
Our study equally applies to both unpolarized and polarized 
scattering processes.

The paper is organized as follows. In Sect.~\ref{sec:preli} we introduce 
our notation. In Sect.~\ref{sec:nlo} we present the result of our general NLO 
calculation of the partonic cross section. The resummation of the 
logarithmically-enhanced terms and the all-order resummation formula
are presented and discussed in Sect.~\ref{sec:resu}.
Our results are briefly summarized in Sect.~\ref{sec:summa}.

\section{Single-particle cross section and notation}
\label{sec:preli}

We consider the inclusive hard-scattering reaction
\begin{equation}
\label{hadpro}
h_1(P_1)+h_2(P_2)\to h_3(P_3)+X \;,
\end{equation}
where the collision of the two hadrons $h_1$ and $h_2$ with momenta $P_1$ and $P_2$, respectively,
produces the hadron $h_3$ with momentum $P_3$ accompanied by
an arbitrary and undetected final state $X$.
According to the QCD factorization theorem the corresponding cross section is given by
\begin{align}
\label{factorization}
E_3\f{d\sigma_{h_3}}{d^3\bP_3}(P_1,P_2,P_3)&=\sum_{a_1,a_2,a_3}\int_0^1 dx_1
\int_0^1 dx_2 \int_0^1 \f{dx_3}{x_3^2} \;
f_{a_1/h_1}(x_1,\mu_F)\, f_{a_2/h_2}(x_2,\mu_F)\, d_{a_3/h_3}(x_3,\mu_f)\nn\\
& \times \;p^0_3 \f{d{\hat \sigma}_{a_1a_2\to a_3}}{d^3\bp_3}(x_1 P_1,x_2
P_2,P_3/x_3;\mu_F,\mu_f) \;,
\end{align}
where the index $a_i \,(i=1,2,3)$ denotes the parton species
($a=q,{\bar q},g$),  $f_{a/h}(x,\mu_F)$ is the parton density of the colliding
hadron evaluated at the
factorization scale $\mu_F$, and $d_{a/H_3}(x,\mu_f)$ is the fragmentation function of the parton $a$ into
the hadron $H_3$ at the factorization scale $\mu_f$ (in general, the
fragmentation scale $\mu_f$ can be different from the scale $\mu_F$ of the parton
densities). We use parton densities and fragmentation functions as defined
in the \ms\ factorization scheme. The last factor, 
$d{\hat \sigma}_{a_1a_2\to a_3}(p_1,p_2,p_3)$, on the right-hand side of
Eq.~(\ref{factorization}) is the inclusive cross section for the partonic subprocess
\begin{equation}
\label{partpro}
a_1(p_1)+a_2(p_2)\to a_3(p_3)+X\,\,,
\end{equation}
which, throughout the paper, is always treated with massless partons (kinematics).

In Eq.~(\ref{factorization}), the partonic (hadronic) Lorentz-invariant phase
space $d^3\bp_3/p^0_3$ ($d^3\bP_3/E_3$) is explicitly denoted in terms of the
energy $p^0_3 \,(E_3)$ and the three-momentum $\bp_3 \,(\bP_3)$ of the `detected'
final-state parton $a_3$ (hadron $h_3$). Other kinematical variables can
equivalently be used. For instance, considering the centre--of--mass frame
of the two colliding partons in the partonic subprocess of Eq.~(\ref{partpro}), 
we have
\beq
\label{etapt}
d^3\bp_3/p^0_3 = d\eta \;d^2{\bf p}_T \;\;, \quad
\quad \eta=\f{1}{2} \ln\f{p_3\cdot p_2}{p_3\cdot p_1} \;\;,
\eeq
where ${\bf p}_T$ is the transverse-momentum of the parton $a_3$ and $\eta$
is its rapidity (the forward region $\eta > 0$ corresponds to the direction of the
parton $a_1$). The kinematics of the partonic subprocess can also be described
by using the customary Mandelstam variables $s, t, u$:
\begin{equation}
\label{defstu}
s=(p_1+p_2)^2~,~~~t=(p_1-p_3)^2~,~~~u=(p_2-p_3)^2~,
\end{equation}
with the phase-space boundaries
\begin{equation}
s \geq 0~,~~~t \leq 0~,~~~u \leq 0~,~~~s+t+u \geq 0 \;\;. 
\end{equation}
Analogous kinematical variables can be introduced for the corresponding
hadronic process in Eq.~(\ref{hadpro}). 
Throughout the paper, hadronic and partonic kinematical variables
are typically denoted by the same symbol, although we use capital letters
for hadronic variables. For instance, $S=(P_1+P_2)^2$ is the square of the
centre--of--mass energy of the hadronic collision and $\bP_{3T}=\bP_T$
is the transverse momentum of the observed hadron $h_3$.

The partonic cross section $d{\hat \sigma}_{a_1a_2\to a_3}$ depends 
on the factorization scales, and it
is computable in QCD perturbation theory 
as power series expansion in the QCD coupling $\as(\mu_R^2)$ ($\mu_R$ denotes
the renormalization scale, and we use the \ms\ renormalization scheme).
The perturbative expansion starts at ${\cal O}(\as^2)$
since the leading order (LO) partonic process corresponds to the $2\to 2$ reaction $a_1a_2\to a_3a_4$. 
Considering the expansion up to the next-to-leading order (NLO), we write
\beeq
\label{pertsig}
\!\!\! d{\hat \sigma}_{a_1a_2\to a_3}(p_1,p_2,p_3;\mu_F,\mu_f)
=\as^2(\mu_R^2)\!\!\!\!\!\!&&\!\!\!\Bigl( \; d{\hat \sigma}_{a_1a_2\to
a_3a_4}^{(0)}(p_1,p_2,p_3) \Bigr.  \\
\!\!&&\!\!+\Bigl. \;\f{\as(\mu_R^2)}{2\pi}
\,d{\hat \sigma}^{(1)}_{a_1a_2\to a_3}(p_1,p_2,p_3;\mu_R,\mu_F,\mu_f)
+{\cal O}(\as^2)\Bigr)\, . \nn
\eeeq
The LO term $d{\hat \sigma}_{a_1a_2\to a_3a_4}^{(0)}$ is directly related 
(see Eq.~(\ref{sigborn}))
to the Born-level scattering amplitude of the partonic reaction  
$a_1a_2\to a_3a_4$. The NLO term $d{\hat \sigma}^{(1)}_{a_1a_2\to a_3}$ is 
known: the contribution of the partonic subprocess with non-identical quarks
was computed in Refs.~\cite{Ellis:1979sj, Aversa:1988fv}, 
and the complete NLO calculation for all
partonic subprocesses was presented in Ref.~\cite{Aversa:1988vb}.

The NLO calculation was carried out in analytical form, and it is presented
\cite{Ellis:1979sj, Aversa:1988fv, Aversa:1988vb} in
terms of the independent kinematical variables $s, v$ and $w$, which are related
to the Mandelstam variables of Eq.~(\ref{defstu}) through the definition
\begin{equation}
\label{defvw}
v \equiv 1+t/s~,~~~~~~~~w \equiv -u/(s+t)\,\,,
\end{equation}
with the corresponding phase-space boundaries
\begin{equation}
s \geq 0~,~~~1 \geq v \geq 0~,~~~1 \geq w \geq 0 \;\;. 
\end{equation}
Using these variables, the partonic cross section in Eqs.~(\ref{factorization}) 
and (\ref{pertsig}) can be written as
\beeq
\label{fixedorder}
p^0_3\f{d{\hat \sigma}}{d^3\bp_3}(p_1,p_2,p_3;\mu_F,\mu_f)=
\f{\as^2(\mu_R^2)}{\pi \,s}&&\!\!\!\!\!\!
\Bigl[ \;\f{1}{v}\f{d{\hat \sigma}^{(0)}(s,v)}{dv}\;\delta(1-w)
\Bigr.  \\
&&\!\!\!\!\!\!+\Bigl. \; \f{\as(\mu_R^2)}{2\pi} \;\f{1}{v \,s}
\;{\cal C}^{(1)}(s,v,w;\mu_R,\mu_F,\mu_f)
+{\cal O}(\as^2)
\Bigr]\,\, , \nn
\eeeq
where the flavour indices are left understood (the term in the square bracket
exactly corresponds to the square-bracket term in Eq.~(10) of 
Ref.~\cite{Aversa:1988vb},
modulo the overall factor $\as^2(\mu_R^2)$).
The first term in the square bracket of Eq.~(\ref{fixedorder}) is the Born-level
contribution,
and the function ${\cal C}^{(1)}$ encodes the NLO corrections.

The Born-level term in Eq.~(\ref{fixedorder}) has a sharp integrable 
singularity at $w=1$. This singularity has a kinematical origin.
Indeed $(1 - w)$ is proportional (see Eq.~(\ref{defvw})) to $s_X= s + t + u$,
which is the invariant mass squared of the QCD radiation (i.e. the unobserved
final-state system $X$ in Eq.~(\ref{partpro})) recoiling against the `observed'
final-state parton $a_3$. At the LO, the system $X$ is formed by a single {\em
massless} parton $a_4(p_4)$ and, therefore, $s_X=p_4^2$ exactly vanishes thus
leading to the factor $\delta(1-w)$ in Eq.~(\ref{fixedorder}). At higher
perturbative orders,
the LO singularity at $w \to 1$ is enhanced by logarithmic terms of the type
$\ln(1-w)$. The enhancement has a dynamical origin, and it is produced by
soft-gluon radiation. Indeed, in the kinematical region where $w \to 1$,
the system $X$ is forced to carry a very small invariant mass, and the
associated production of hard QCD radiation is strongly suppressed.
The associated production of soft QCD radiation is instead allowed and, due to
the soft-gluon bremsstrahlung spectrum, it generates large logarithmic
corrections.

The presence of logarithmically-enhanced terms is evident from the known NLO
result.
The structure of the NLO term ${\cal C}^{(1)}$ in Eq.~(\ref{fixedorder})
is customarily written (see, e.g., Eqs.~(10) and (22) in 
Ref.~\cite{Aversa:1988vb}) in the following form:
\beeq
\label{C1}
{\cal C}^{(1)}(s,v,w;\mu_R,\mu_F,\mu_f)&=&{\cal C}_3(v)\left(\f{\ln(1-w)}{1-w}\right)_+
+{\cal C}_2(v;s,\mu_F,\mu_f) \left(\f{1}{1-w}\right)_+ \nn \\
&+&{\cal C}_1(v;s,\mu_R,\mu_F,\mu_f)\, \delta(1-w)
+ {\cal C}_0(1-w, v;s,\mu_R,\mu_F,\mu_f)\, .
\eeeq
The last term on the right-hand side is a non-singular function of $w$ in the
limit $w \to 1$, namely, ${\cal C}_0(1-w, v)={\cal O}\left((1-w)^0\right)$
(see Refs.~\cite{Ellis:1979sj, Aversa:1988fv} for explicit expressions in
analytic form). The functions ${\cal C}_3$, ${\cal C}_2$ and ${\cal C}_1$
do not depend on $w$, and they multiply functions of $w$ that are
singular (and logarithmically-enhanced)
at $w \to 1$. These singular functions are expressed by
$\delta(1-w)$ and customary `plus-distributions', $[(\ln^k(1-w))/(1-w)]_+\,$,
defined over the range $1 \geq w \geq 0$.

In this paper we deal with the perturbative QCD contributions beyond the
NLO, in the kinematical region where $w \to 1$ or, more {\em generally},
$s_X \to 0$. This region is usually referred to as the region of {\em partonic}
threshold, since the partonic process in Eq.~(\ref{partpro}) approaches the
near-elastic limit. The observed parton $a_3$ is produced with the maximal
energy that is kinematically allowed by momentum conservation, and the recoiling
partonic system $X$ has the minimal invariant mass.
We are interested in the near-threshold behaviour of the
partonic cross section, and we compute the higher-order contributions that
dominate near the partonic threshold. Before considering the higher-order terms,
in Sect.~\ref{sec:nlo} we focus on the behaviour of the NLO cross section, and
we present the results of our independent NLO calculation in the kinematical
region close to the partonic threshold. Our NLO results are obtained and
expressed in a form that is suitable (and necessary) for the all-order treatment
and resummation of the logarithmically-enhanced QCD corrections. 

The discussion in this section has been limited to the case in which
the inclusive hard-scattering reaction in Eq.~(\ref{hadpro}) is
unpolarized. The same discussion applies to polarized process in which one or
more of the three hadrons $h_1, h_2$ and $h_3$ have definite states of spin
polarizations. The only difference between the unpolarized and polarized cases is
that the parton densities, the fragmentation function and the partonic cross
section in the factorization formula (\ref{factorization}) have to be replaced by
the corresponding spin-polarized quantities. The structure of the threshold
behaviour of the polarized partonic cross section is completely analogous to that
of Eqs.~(\ref{fixedorder}) and (\ref{C1})
(see, e.g., Ref.~\cite{de Florian:2007ty} and references therein)
In the following sections, we continue our discussion by explicitly considering
the unpolarized case. Our results equally apply to both unpolarized and polarized
cross sections. At the end of Sect.~\ref{sec:resu} 
(just before Sect.~\ref{sec:rapint}),
we briefly comment on soft-gluon resummation for the polarized case, 
and we summarize the technical
differences between unpolarized and polarized scattering processes.

\section{NLO results near partonic threshold}
\label{sec:nlo}



In the near-threshold region, the NLO partonic cross section of 
Eq.~(\ref{fixedorder}) is controlled by the functions 
${\cal C}_3, {\cal C}_2$ and  ${\cal C}_1$ in Eq.~(\ref{C1}) and, more
precisely, each of these functions depends on the various flavour channels that
contribute to the partonic reaction $a_1a_2 \to a_3 a_4$. The functions
${\cal C}_{i, a_1a_2 \to a_3 a_4}(v)$ with $i=1,2,3$ are all reported in Sect.~3
of Ref.~\cite{Aversa:1988vb}. The corresponding analytic expressions have a
rather involved dependence (especially for ${\cal C}_1$) on $v$, colour factors
and the flavour channel.

We have performed an independent calculation of the NLO cross section near
threshold. We have computed the three coefficients of the logarithmic
expansion in Eq.~(\ref{C1}), including the coefficient ${\cal C}_1$ that
controls the term proportional to 
$\delta(1-w)$. The final result is presented in this section, and it has a
rather compact form. More importantly, it embodies an amount of 
{\em process-independent} information that cannot be extracted (or, say, it
is difficult to be extracted) from the results of Ref.~\cite{Aversa:1988vb}.
In particular, our calculation and the ensuing result keep
explicitly under control {\em colour correlation} effects
that are a typical and general feature of soft-gluon radiation from 
$2 \to 2$ parton scattering processes.
The knowledge of these colour correlation terms
is essential (see Sect.~\ref{sec:resu}) to compute logarithmically-enhanced
contributions beyond the NLO.

At the NLO, the parton cross section receives contributions from two types of
partonic processes. The elastic process
\begin{equation}
\label{elpro}
a_1(p_1)+a_2(p_2)\to a_3(p_3)+ a_4(p_4) \,\,,
\end{equation}
which has to be evaluated with one-loop virtual corrections, and the inelastic
process in Eq.~(\ref{partpro}) with real emission of $X=\{ 2~{\rm partons} \}$, which
is evaluated at the tree level.
Virtual and real contributions are separately divergent, and we use dimensional
regularization with $d=4-2\ep$ space-time dimensions to deal with both
ultraviolet and infrared (IR) divergences. The elastic process 
contributes only to the term proportional to $\delta(1-w)$ in Eq.~(\ref{C1}),
and its contribution is directly proportional to the (ultraviolet)
renormalized one-loop scattering amplitude of the four-parton process.
In the threshold region $w\to 1$, the five-parton inelastic process gives
dominant contributions only from two kinematical configurations of the system
$X=\{ 2~{\rm partons} \}$: either one of the two partons is soft or both
partons are collinear. We treat these two configurations by using soft and
collinear factorization formulae (in colour space) \cite{Catani:1996vz} 
of the scattering amplitudes, and we perform the phase-space integration.
This real emission term is finally combined with the collinear-divergent
counterterms necessary to define the NLO parton densities and fragmentation
function and with the virtual correction from the four-parton elastic process.
The final result, which is IR finite, has a factorized structure:
it is given in terms of  
flavour and colour-space factors that acts on the scattering amplitude
of the four-parton elastic process.

To present the result of our NLO calculation in its factorized form,  
we need to briefly recall the representation of the four-parton scattering
amplitude in the colour-space notation \cite{Catani:1996vz, Catani:1998bh}.
The {\em all-loop} QCD amplitude ${\cal M}$ of the scattering process 
in Eq.~(\ref{elpro}) is written as
\begin{equation}
\label{ampl}
|{\cal M}_{a_1a_2a_3a_4}\rangle= 
\as(\mu_R^2)\left[|{\cal M}^{(0)}_{a_1a_2a_3a_4}\rangle
+\sum_{n=1}^\infty\left(\f{\as(\mu_R^2)}{2\pi}\right)^n|{\cal
M}^{(n)}_{a_1a_2a_3a_4}(\mu_R)\rangle\right] \;\;,
\end{equation}
where ${\cal M}^{(0)}$ is the Born-level contribution, ${\cal M}^{(n)}$
is the 
contribution at the $n$-loop level, and we always consider the renormalized 
(in the \ms\ scheme) amplitude. The remaining IR divergences are regularized 
in $d=4-2\ep$ space-time dimensions by using the customary scheme of
conventional dimensional regularization (CDR) \cite{Ellis:1985er}.
The subscript `$a_1a_2a_3a_4$' refers
to the flavour of the four partons, while the dependence on the parton momenta 
$p_i$ $(i=1,\dots,4)$ is not explicitly denoted. Note, however, that the
elastic $2 \to 2$ process is evaluated exactly at the partonic threshold
(i.e. with $s+t+u=0$), and
momentum conservation ($p_1+p_2=p_3+p_4$) implies that ${\cal M}$ only depends
on two kinematical variables (e.g., it depends on $s$ and $v$).

The colour indices $c_i$ of the partons are embodied  
\cite{Catani:1996vz, Catani:1998bh} in the `ket' notation, through the
definition
\beq
{\cal M}_{a_1a_2a_3a_4}^{c_1c_2c_3c_4} \equiv 
\langle \,c_1c_2c_3c_4 |{\cal M}_{a_1a_2a_3a_4}\rangle \;\;,
\eeq
so that $| \cdots \rangle$ is an abstract vector in colour space, and 
$\langle \cdots |$ is its complex-conjugate vector. Gluon radiation from the
parton with momentum $p_i$ is described by the colour-charge matrix 
$(\T_i)^c$ ($c$ is the colour index of the radiated gluon) 
and colour
conservation implies
\beq
\label{colcons}
\sum_{i=1}^4 \; \T_i \;|{\cal M}_{a_1a_2a_3a_4}\rangle = 0 \;\;.
\eeq
Note that according to this notation the colour flow is treated as `outgoing',
so that $\T_3$ and $\T_4$ are the colour charges of the partons $a_3$ and
$a_4$, while $\T_1$ and $\T_2$ are the colour charges of the anti-partons 
${\overline a}_1$ and ${\overline a}_2$.
The colour-charge algebra for the product 
$(\T_i)^c (\T_j)^c \equiv \T_i \cdot \T_j$ gives
\beq
\T_i^2 = C_{a_i} \;, \quad \quad \quad \quad C_g=C_A \;, \;\;\; C_q=C_{\bar q}=C_F \;,
\eeq
where $C_a$ is the Casimir factor and, in $SU(N_c)$ QCD, we have $C_a=N_c$ if $a$ 
is a gluon and $C_a=C_F=(N_c^2-1)/(2N_c)$ if $a$ is a quark or an antiquark.
Thus, $\T_i^2$ is a $c$-number term or, more precisely, a multiple of the unit
matrix in colour space.
Non-trivial {\em colour correlations} are produced by the quadratic operators
$\T_i \cdot \T_j= \T_j \cdot \T_i$ with $i\neq j$. These are six different
operators, but, due to colour conservation (i.e. Eq.~(\ref{colcons})), only {\em
two} of them lead to colour correlations that are linearly independent 
(see the Appendix~A of Ref.~\cite{Catani:1996vz}). Two linearly independent
operators are $\T_1 \cdot \T_3$ and $\T_2 \cdot \T_3$. Different choices of
pairs (e.g., the pair $\T_1 \cdot \T_2$ and $\T_1 \cdot \T_3$)
of independent operators are feasible and physically equivalent.
For instance, by analogy with the Mandelstam kinematical variables of the $2
\to 2$ parton scattering, we can use \cite{Dokshitzer:2005ig} 
the $s$-~, $t$-~ and 
$u$-channel colour-correlation operators ${\bf T}_s^2$, ${\bf T}_t^2$ and 
${\bf T}_u^2$,
\beeq
\label{corcharge}
{\bf T}_s^2 &=& \left({\bf T}_1+{\bf T}_2 \right)^2 = 
\left({\bf T}_3+{\bf T}_4 \right)^2\;\;, \nn\\
{\bf T}_t^2 &=& \left({\bf T}_1+{\bf T}_3 \right)^2 = 
\left({\bf T}_2+{\bf T}_4 \right)^2\;\;, \\
{\bf T}_u^2 &=& \left({\bf T}_2+{\bf T}_3 \right)^2 = 
\left({\bf T}_1+{\bf T}_4 \right)^2\;\;, \nn
\eeeq
which are linearly related by colour conservation:
\beq
\label{sumcorr}
{\bf T}_s^2 + {\bf T}_t^2 + {\bf T}_u^2 = \sum_{i=4}^4 {\bf T}_i^2 = 
\sum_{i=4}^4  C_{a_i}\;\;.
\eeq

The LO cross section in Eq.~(\ref{fixedorder}) depends on the square of the
Born-level scattering amplitude $|{\cal M}^{(0)}\rangle$:
\beq
\label{sigborn}
\f{d{\hat \sigma}^{(0)}_{a_1a_2\to a_3a_4}(s,v)}{dv} = \f{1}{N^{(in)}_{a_1a_2}}
\;\f{1}{16 \pi \,s} \; |{\cal M}^{(0)}_{a_1a_2a_3a_4}|^2 \;\;,
\eeq
where $|{\cal M}^{(0)}|^2 = \langle {\cal M}^{(0)}\;|\;{\cal M}^{(0)} \rangle$
and the factor $N^{(in)}_{a_1a_2}=4\,n_c(a_1)\,n_c(a_2)$ comes from the average
over the spins and colours ($n_c(q)=n_c({\bar q})=N_c, n_c(g)=N_c^2-1$) of the
initial-state partons $a_1$ and $a_2$.

The Born-level and one-loop ($|{\cal M}^{(1)}\rangle$) scattering amplitudes
of the partonic reaction $a_1a_2\to a_3a_4$ are known 
\cite{Bern:1991aq, Kunszt:1993sd}.
The one-loop scattering amplitude includes IR-divergent terms that have a 
process-independent (universal) structure \cite{Giele:1991vf, Catani:1998bh}.
The NLO contribution ${\cal C}^{(1)}$ in Eq.~(\ref{fixedorder}) depends on the
IR-finite part ${\cal M}^{(1)\,{\rm fin}}$ of the one-loop scattering amplitude.
The IR-finite part is obtained through the
factorization formula \cite{Catani:1998bh}
\begin{equation}
\label{m1fin}
|{\cal M}^{(1)}\rangle={\bom I}_{\rm sing}^{(1)} \;|{\cal M}^{(0)}\rangle+
|{\cal M}^{(1)\,{\rm fin}}\rangle\; ,
\end{equation}
where the colour operator ${\bom I}^{(1)}_{\rm sing}$ embodies 
the one-loop IR divergence in the form of double and single poles 
($1/\ep^2$ and $1/\ep$), while ${\cal M}^{(1)\,{\rm fin}}$ is finite as 
$\ep\to 0$. To specify the expression of ${\cal M}^{(1)\,{\rm fin}}$ in an
unambiguous way, the contributions of ${\cal O}(\ep^0)$ that are included in
${\bom I}^{(1)}_{\rm sing}$ must be explicitly defined.  We use the expression
\begin{equation}
\label{I1sing}
{\bom I}^{(1)}_{\rm sing}= \frac{1}{2}\frac{1}{\Gamma(1-\ep)}\left[
\f{1}{\ep^2}
\sum_{\substack{i,j=1 \\ i \,\neq\, j}}^4
\T_i\cdot\T_j
\left(\frac{4\pi\mu_R^2\,e^{-i\lambda_{ij}\pi}}{2 p_i \cdot p_j}\right)^{\epsilon}
-\f{1}{\ep}
\sum_{i=1}^4
\gamma_{a_i}\left(\f{4\pi\mu_R^2\, s}{u\,t}\right)^\ep\right]
\;\;,
\end{equation}
where 
$e^{-i\lambda_{ij}\pi}$ is the unitarity phase factor ($\lambda_{ij}=-1$ if $i$
and $j$ are both incoming or outgoing partons and $\lambda_{ij}=0$ otherwise),
and the flavour dependent coefficients $\gamma_{a}$ are
($n_F$ is the number of flavours of massless quarks)
\begin{align}
\label{gamcoef}
&\gamma_q=\gamma_{\bar q}=\f{3}{2}C_F~~,~~~~~~\gamma_g=\f{11}{6}C_A-\f{1}{3}n_F
\;\; .
\end{align}
Note that the operator ${\bom I}^{(1)}_{\rm sing}$ in Eq.~(\ref{I1sing}) differs
from the operator ${\bom I}^{(1)}$ used in Ref.~\cite{Catani:1998bh}: the
difference is IR finite, and it is due to terms of ${\cal O}(\ep^0)$ that are
proportional to the coefficients $\gamma_{a_i}$. The effect of this difference is
absorbed in $|{\cal M}^{(1)\,{\rm fin}}\rangle$.

The term  ${\cal C}^{(1)}$ in Eq.~(\ref{fixedorder}) is IR finite, and the final
result of our NLO calculation is expressed by the following colour-space
factorization formula:
\begin{equation}
\label{nlofactform}
16 \pi \,N^{(in)} \;{\cal C}^{(1)}=\langle {\cal M}^{(0)}|\,{\bom C}^{(1)}|{\cal M}^{(0)}\rangle
+\left(\langle {\cal M}^{(0)}|{\cal M}^{(1)\,{\rm fin}}\rangle+{\rm c.c.}\right)\delta(1-w)+{\cal O}\left((1-w)^0\right)\, ,
\end{equation}
where ${\rm c.c.}$ stands for complex conjugate, and the flavour indices are left
understood. The function ${\bom C}^{(1)}$ (reinserting the dependence on flavour
indices and kinematical variables, we actually have
${\bom C}^{(1)} \to{\bom C}^{(1)}_{a_1a_2a_3a_4}(s,v,w;\mu_R, \mu_F, \mu_f)$)
has the form of a colour-space operator. We find the result:
\begin{align}
\label{C1ris}
{\bom C}^{(1)}_{a_1a_2a_3a_4}(s,v,w;\mu_R,\mu_F,\mu_f)
    &= 2\plusd{\ln(1-w)}{1-w} \bigg[\, 2\sum_{i=1}^3\T_i^2 - \T_4^2 \,\bigg] \notag\\
    &- \plusd{1}{1-w} \bigg[\, 2\sum_{i=1}^3\T_i^2 \left(\ln\frac{1-v}{v} + \ln\frac{\mu_{Fi}^2}{s}\right) - 2\,\T_4^2\ln(1-v) \notag\\
        &+ \gamma_{a_4} + 8\,\Big(\T_1\cdot \T_3\ln(1-v) + \T_2\cdot\T_3\,\ln v\Big) \,\bigg] \notag\\
    &+ \delta(1-w) \Bigg\{\f{\pi^2}{2}\left(\T_1^2+\T_2^2+3\T_3^2-\f{4}{3}\T_4^2\right)\nn\\
    &- 2\T_3^2\ln v\,\ln\f{\mu_f^2}{s}+2\T_2^2\ln\f{1-v}{v}\ln\f{\mu_F^2}{s}-\sum_{i=1}^3\gamma_{a_i}\ln\frac{\mu_{Fi}^2}{s\,v(1-v)}\nn\\
&+\ln v\ln (1-v) \left(\T_4^2-\T_1^2-\T_2^2-\T_3^2\right)\nn\\
&+\ln^2(1-v) \left(\T_1^2+\T_3^2-\T_4^2\right)+\ln^2 v\,(\T_2^2+\T_3^2)+\gamma_{a_4} \ln (1-v)\nn\\
&+\T_1\cdot \T_3 \Bigl(2\pi^2+2\ln(1-v)\left(\ln(1-v)-2\ln v\right)\Bigr)\nn\\
&+\T_2\cdot \T_3 \Bigl(2\pi^2+2\ln v\left(2\ln(1-v)-3\ln v\right)\Bigr)+K_{a_4}
\Bigg\}\,,
\end{align}
where we have defined
\begin{equation}
\label{facscales}
\mu_{F1}=\mu_{F2}=\mu_F~~,~~~~~~~\mu_{F3}=\mu_f\; ,
\end{equation}
and the flavour dependent coefficients $K_a$ are given by
\begin{equation}
\label{Kappa}
K_q=K_{\bar
q}=\left(\f{7}{2}-\f{\pi^2}{6}\right)C_F~~,~~~~~K_g=\left(\f{67}{18}-\f{\pi^2}{6}\right)C_A-\f{5}{9}\,n_F\; .
\end{equation}

The result in Eq.~(\ref{C1ris}) contains terms that are proportional to
plus-distributions of $w$ (the action of these terms onto the
Born-level scattering amplitude as in Eq.~(\ref{nlofactform})
directly gives the coefficients 
${\cal C}_3$ and ${\cal C}_2$ in Eq.~(\ref{C1})) and a term that is proportional
to $\delta(1-w)$. The sum of the latter term and the analogous term (which is
proportional to ${\cal M}^{(1) \,{\rm fin}}$) on the right-hand side of 
Eq.~(\ref{nlofactform}) gives the function ${\cal C}_1$ in Eq.~(\ref{C1})
(note that a change in the definition of ${\cal M}^{(1) \,{\rm fin}}$
would be compensated by a corresponding change in ${\bom C}^{(1)}$, so that the
total NLO result in Eqs.~(\ref{C1}) and (\ref{nlofactform}) is unchanged).

All the contributions to the NLO colour-space function ${\bom C}^{(1)}$ in 
Eq.~(\ref{C1ris}) have a definite physical origin. The terms that are
proportional to the colour charges $\T_i$ are due to radiation 
(either collinear or at wide angles)
of soft gluons. In particular, the coefficients of $(1/(1-w))_+$ and 
$\delta(1-w)$ depend on colour correlation operators. In Eq.~(\ref{C1ris}),
we have used the two linearly independent operators $\T_1\cdot \T_3$ and
$\T_2\cdot \T_3$ to explicitly present the colour correlation contributions.
The terms that are proportional to the flavour-dependent coefficients
$\gamma_a$ and $K_a$ have a collinear (and non-soft) origin.
In particular, we recall (see Eq.~(C.13) in Appendix~C of 
Ref.~\cite{Catani:1996vz})
that $K_a$ is related to the $(d-4)$-dimensional part (i.e. the terms of
${\cal O}(\ep)$) of the LO collinear splitting functions. We also remark and
recall (see Eq.~(7.28) and related comments in Ref.~\cite{Catani:1996vz}) that
the gluonic coefficient $K_g$ in Eq.~(\ref{Kappa})
is exactly equal to the coefficient $K$ (see Eqs.~(\ref{acoef}) and (\ref{cmw}))
that controls the intensity of soft-gluon radiation at ${\cal O}(\as^2)$.

In the case of the four-parton scattering $a_1a_2\to a_3a_4$, 
our process-independent NLO results can be checked by comparison with the NLO
results of Ref.~\cite{Aversa:1988vb}.
Using Eqs.~(\ref{m1fin}), (\ref{nlofactform}) and (\ref{C1ris}) and 
the one-loop virtual contributions from Ref.~\cite{Ellis:1985er}, we have
verified that we correctly reproduce the results of Ref.~\cite{Aversa:1988vb}
for the NLO coefficient ${\cal C}^{(1)}$ of the various partonic channels
(note that the expressions of Ref.~\cite{Aversa:1988vb} have to be converted to
the \ms\ factorization scheme, since they explicitly refer to a different
factorization scheme).

An additional check can be carried out by considering the case in which the
parton $a_3$ is replaced by a photon. In this case $\T_3=0$, and the colour
algebra becomes trivial (the colour correlation terms 
$\T_1\cdot \T_3$ and $\T_2\cdot \T_3$ vanish). Using the one-loop virtual
contribution for the process $g q \to\gamma q$ \cite{Aurenche:1986ff}
and its crossing-related channels,
we have explicitly verified that the results in 
Eqs.~(\ref{m1fin}), (\ref{nlofactform}) and (\ref{C1ris})
correctly reproduces the NLO coefficient of the cross section
for prompt-photon production 
\cite{Aurenche:1983ws, Gordon:1993qc}.

\section{All-order soft-gluon resummation}
\label{sec:resu}

In the near-threshold region $w \to 1$, the singular behaviour of the NLO 
partonic cross section is further enhanced at higher perturbative orders. 
Radiation of soft
and collinear partons can produce (at most) two additional powers of
$\ln (1-w)$ for each additional power of $\as$. A reliable evaluation of the
partonic cross section in the near-threshold region requires the computation and,
possibly, the all-order resummation of these large logarithmic contributions.

Note that we are considering the partonic (and not the hadronic) cross section 
in its near-threshold region. The available partonic phase space is smaller than
the hadronic phase space. Therefore, if the hadronic process in Eq.~(\ref{hadpro})
is studied in kinematical configurations close to its threshold
(roughly speaking, the region where $P_{3T} \simeq {\sqrt S}/2$),
the partonic process in Eq.~(\ref{partpro})
is also kinematically forced toward its threshold.
In these kinematical configurations, the behaviour of the hadronic
cross section is certainly dominated by the large logarithmic contributions.
Nonetheless, as is well known, these partonic logarithmic contributions
typically (see, e.g., Ref.~\cite{deFlorian:2005yj}) 
give the bulk of the radiative corrections to the hadronic process also in
kinematical configurations that are not 
close to the hadronic threshold.
This effect is due to the convolution structure  with the parton densities
and the fragmentation function according to Eq.~(\ref{factorization}).
Roughly speaking, the partonic threshold corresponds to the region where
$p_{3T} \simeq {\sqrt s}/2$, which can be rewritten in terms of hadronic
variables
($p_{3T}=P_{3T}/x_3 \;, s= x_1x_2 S$ as in Eq.~(\ref{factorization}))
and it translates into the region where
$P_{3T} \simeq x_3{\sqrt {x_1x_2}}{\sqrt S}/2$ . Since the typical average 
values of momentum fractions $x_i$ ($i=1,2,3$) that mostly contribute to
Eq.~(\ref{factorization})
are small (parton densities and fragmentation functions are indeed strongly
suppressed at large values of $x$), the partonic threshold region
$P_{3T} \simeq x_3{\sqrt {x_1x_2}}{\sqrt S}/2$ can give the dominant contribution
to the hadronic cross section even if $P_{3T} \ll {\sqrt S}/2$, namely,
in kinematical configurations that are far from the hadronic threshold.

The three independent kinematical variables $\{ \,s, v, w\,\}$ (which are
customarily used to present the NLO results) are not particularly suitable for
an all-order treatment near threshold, because of their degree of asymmetry under
the exchange $u \leftrightarrow t$. The all-order treatment of the terms 
$\ln^n (1-w)$ unavoidably produces an asymmetry with respect to 
$u \leftrightarrow t$ (see Eq.~(\ref{defvw})). In practical applications of
resummation, this feature can lead to
(quantitatively) non-negligible and unphysical asymmetries in the angular
(rapidity) distribution of the produced hadron $h_3$.
We note that this asymmetry effect is formally suppressed by powers of $(1-w)$
{\em only} after the complete resummation of the entire perturbative series of
logarithmic terms to all orders in $\as$. Any feasible resummed calculations
involve the truncation of the all-order series to some level of logarithmic
accuracy and, in this case, the asymmetry effect is suppressed only by
subleading (but still singular) logarithmic contributions (see
Eqs.~(\ref{plusnlo}) and (\ref{plusall})).

We introduce the three independent kinematical variables
$\{ \,\xw, r, p_T^2\,\}$ that are defined by
\begin{equation}
\label{defxrpt}
\xw=-\f{u+t}{s}~,~~~~r=\f{u}{t}~,~~~~p_T^2=\f{ut}{s} \;,
\end{equation}
with the corresponding phase-space boundaries
\begin{equation}
1 \geq \xw \geq 0~,~~~~r \geq 0~,~~~~p_T^2 \geq 0 \;.
\end{equation}
The variable $p_T$ is the transverse momentum of the observed parton $a_3$
(see Eq.~(\ref{etapt})). In the centre--of--mass frame of the partonic collision
in Eq.~(\ref{partpro}), the variable $\xw= 2\,p_3^0/{\sqrt s}$ is the energy
fraction of the parton $a_3$ and 
$r=(1 +\cos \theta^*_{13})/(1 -\cos \theta^*_{13})$ is related to its scattering
angle $\theta^*_{13}$. The relation with the transverse momentum and rapidity of
the parton $a_3$ (see Eq.~(\ref{etapt})) is
\beq
\label{kineta}
\xw=\f{2\,p_T}{{\sqrt s}} \cosh \eta~,~~~~r=e^{2 \eta} \;.
\eeq
In terms of the kinematical variables in Eq.~(\ref{defxrpt}), the near-threshold
limit $s_X = s+t+u \to 0$ corresponds to the region where $\xw \to 1$, at fixed
values of $p_T$ and $r$. Therefore, the threshold variable is $\xw$, and it is
symmetric with respect to the exchange $u \leftrightarrow t$.

The change of variables 
$\{ \,s, v, w\,\} \leftrightarrow \{ \,\xw, r, p_T^2\,\}$
can be straightforwardly applied to any smooth functions of these variables.
Singular (plus) distributions require a slightly more careful treatment, because
of the presence of contact terms at the endpoints $w=1$ and $\xw =1$. We have
\beeq
\delta(1- \xw) &=& \f{1}{v} \,\delta(1- w) \;\;, \nn \\
\label{plusnlo}
\left[ \f{1}{1-\xw} \right]_+ &=& \f{1}{v}\, \left\{
\;\left[ \f{1}{1-w} \right]_+ \,+ \,\delta(1- w) \,\ln v
\, \right\}
\;\;, \\
\left[ \f{\ln(1-\xw)}{1-\xw} \right]_+ &=& \f{1}{v}\,  \left\{
\;\left[ \f{\ln(1-w)}{1-w} \right]_+ \,+ \left[ \f{1}{1-w} \right]_+ \,\ln v
\,+\f{1}{2} \;\delta(1- w) \,\ln^2 v
\; \right\}\;\;, \nn 
\eeeq
and, more generally,
\beeq
\label{plusall}
\left[ \f{\ln^n (1-\xw)}{1-\xw} \right]_+
&=& \f{1}{v}\,  \left\{
\; \left[ \f{1}{1-w} \left(\,\ln(1-w) + \ln v\,\right)^n \right]_+ 
\,+\;\delta(1- w) \;\f{\ln^{n+1} v}{(n+1)!} 
\; \right\} \nn \\
&=& \f{1}{v}\,  \left\{ \;
\sum_{k=0}^n \f{n!\, \ln^{n-k} v}{k! (n-k)!} 
\; \left[ \f{\ln^k(1-w)}{1-w} \right]_+ 
\,+\;\delta(1- w) \;\f{\ln^{n+1} v}{(n+1)!} 
\; \right\}\;\;.
\eeeq
Using Eq.~(\ref{plusnlo}), the change of variables 
$\{ \,s, v, w\,\} \leftrightarrow \{ \,\xw, r, p_T^2\,\}$
can be applied to the NLO results in Eqs.~(\ref{nlofactform}) and (\ref{C1ris})
and to the complete NLO cross section in Eqs.~(\ref{fixedorder}) and (\ref{C1}).
Note that Eqs.~(\ref{plusnlo}) and (\ref{plusall}) explicitly illustrate 
the previous discussion of the angular ($ u \leftrightarrow t$) asymmetry
effect that arises by using the threshold variable $w$. Indeed, the logarithmic
distribution $\left[ \ln^n (1-\xw)/(1-\xw) \right]_+$ is symmetric with respect
to the exchange $ u \leftrightarrow t$ and, using the variable $w$, this
symmetry is recovered {\em only} throughout the inclusion of many more
subleading (i.e., with $k < n$) logarithmic distributions
$\left[ \ln^k (1-w)/(1-w) \right]_+$, as shown by Eq.~(\ref{plusall}).

Using the kinematical variables in Eq.~(\ref{defxrpt}),
we write the all-order partonic cross section in Eqs.~(\ref{factorization})
and (\ref{pertsig}) in the following form:
\begin{equation}
\label{Sigmadef}
p_3^0\f{d{\hat \sigma}_{a_1a_2\to a_3}}{d^3{\bf
p}_3}=\f{1}{s}\,\sigma^{(0)}_{a_1a_2\to a_3a_4}(r,p_T^2) \;\Sigma_{a_1a_2\to
a_3}(\xw,r;p_T^2,\mu_F,\mu_f) \;\;,
\end{equation}
where the Born-level cross section $\sigma^{(0)}$ is
\begin{equation}
\label{sigborn1}
\sigma^{(0)}_{a_1a_2\to a_3a_4}(r,p_T^2)\equiv
\f{|{\overline {{\cal M}^{(0)}_{a_1a_2 a_3a_4}}}|^2}{16\pi^2 s} \;\;,
\end{equation}
and $|{\overline {{\cal M}^{(0)}}|^2}$ denotes the average of 
$|{\cal M}^{(0)}|^2$ over the spins and colours of the initial-state partons
$a_1$ and $a_2$. The QCD radiative corrections are embodied in the function  
$\Sigma_{a_1a_2\to a_3}$,
\beq
\label{sigmaexp}
\Sigma_{a_1a_2\to a_3}(\xw,r;p_T^2,\mu_F,\mu_f) = 
\as^2(\mu_R^2) \!\left[ \delta(1-\xw) + \sum_{n=1}^{+\infty}
\left(\f{\as(\mu_R^2)}{2 \pi}\right)^n
\Sigma_{a_1a_2\to a_3}^{(n)}(\xw,r;p_T^2,\mu_R,\mu_F,\mu_f)
\, \right] .
\eeq
Note that the LO factor $\as^2(\mu_R^2)$ is included in the definition 
(overall normalization) of $\Sigma$ and, therefore, the radiative function
$\Sigma$ is renormalization group invariant (i.e., the explicit dependence on
$\mu_R$ appears only by expanding $\Sigma$ in powers of $\as(\mu_R^2)$, as in
Eq.~(\ref{sigmaexp})).
We also introduce the definition of the Mellin space $N$-moments $\Sigma_N$
of the function $\Sigma(\xw)$,
\begin{equation}
\label{Nmom}
\Sigma_{a_1a_2\to a_3,N}(r;p_T^2,\mu_F,\mu_f) \equiv
\int_0^1 d\xw \;\xw^{N-1} \;\Sigma_{a_1a_2\to a_3}(\xw,r;p_T^2,\mu_F,\mu_f)
\;\;.
\end{equation}
The $N$ moments are obtained by performing the Mellin transformation 
with respect to the variable $\xw$, at fixed values of  $r$ and $p_T^2$ 
(the hard scale of the partonic process is related to $p_T^2$ rather than to
$s$).

The relations in Eqs.~(\ref{Sigmadef})--(\ref{Nmom}) are simply definitions that
fix our notation. These definitions do not involve any approximations related to
the near-threshold region. The near-threshold limit $\xw\to 1$ corresponds to
the limit $N \to \infty$ in Mellin space.  The $N$ moment of the singular
plus-distribution $\left[ \ln^k (1-\xw)/(1-\xw) \right]_+$
gives  $\ln^{k+1}N$ plus additional subleading logarithms of $N$. The evaluation
(and resummation) of terms with singular distributions of $\xw$ (or $w$)
corresponds to the evaluation (and resummation) of terms with powers of $\ln N$
in Mellin space.

Soft-gluon resummation of near-threshold contributions to single-hadron 
inclusive hadroproduction is studied in Ref.~\cite{deFlorian:2005yj}.
The NLL analysis of Ref.~\cite{deFlorian:2005yj}
deals with the $p_T$ distribution after integration over the rapidity of the
observed hadron. Soft-gluon resummation at fixed rapidity has been examined for
the single-inclusive distribution of a heavy quark 
\cite{Laenen:1998qw, Ahrens:2011mw}
and for the direct component of the cross section in prompt-photon production 
\cite{Laenen:1998qw, Sterman:2000pt, Becher:2009th}.
Soft-gluon resummation for single-hadron production at fixed rapidity requires a
detailed treatment of massless-parton (light-hadron) fragmentation.
Beyond the LL accuracy, fragmentation is not an independent subprocess, since it
is tangled  up with the colour flow dynamics of the entire hard scattering.
Fragmentation in multiparton hard-scattering processes is included in the BCMN
formalism \cite{Bonciani:2003nt},
which we follow and explicitly apply to perform 
soft-gluon resummation for single-hadron inclusive production in hadron
collisions.

We perform resummation in Mellin space \cite{Sterman:1986aj, Catani:1989ne}.
Neglecting contributions of ${\cal O}(1/N)$ that are subdominant in the
near-threshold limit, we write the $N$ moments $\Sigma_N$ of the radiative
function in Eqs.~(\ref{Sigmadef}) and (\ref{Nmom})
in the following form:
\begin{equation}
\label{Nsigmares}
\Sigma_{a_1a_2\to a_3,N}(r;p_T^2,\mu_F,\mu_f) = 
\Sigma_{a_1a_2\to a_3 a_4,N}^{\,{\rm res}}(r;p_T^2,\mu_F,\mu_f) + {\cal O}(1/N) \;\;,
\end{equation}
where $\Sigma_{N}^{\,{\rm res}}$ includes the {\em all-order} resummation of
the $\ln N$ terms (some corrections of ${\cal O}(1/N)$ can also be included in
$\Sigma_{N}^{\,{\rm res}}$). In our resummation treatment, the factorization
scales $\mu_F$ and $\mu_f$ do not play any specific role. The dependence on the
factorization scales and on the renormalization scale $\mu_R$ is treated as in
customary perturbative calculations at fixed order (though the $\ln N$ terms
that enter this dependence are resummed to all orders in $\as$) and, eventually,
the values of $\mu_F,\mu_f$ and $\mu_R$ have to be set to some scale of the
order of $P_T=P_{3T}$, the transverse momentum of the observed hadron.

The all-order expression of $\Sigma_{N}^{\,{\rm res}}$ is obtained by using 
the techniques of Ref.~\cite{Bonciani:2003nt}, which treat soft-gluon
resummation in quite general terms. The BCMN resummation formulae 
\cite{Bonciani:2003nt} apply to arbitrary multiparton hard-scattering processes
and to
general observables that are sensitive to soft-gluon radiation
(the observable should fulfil kinematical properties that are specified in 
Ref.~\cite{Bonciani:2003nt}). The dependence on the specific observable is
parametrized by a Sudakov weight $u(q)$, which is a purely kinematical function.
As discussed in the final part of Ref.~\cite{Bonciani:2003nt}, in our case of
single-particle inclusive production near threshold, the Sudakov weight
is simply $u(q)=\exp \{-N (q\cdot p_4)/(p_1 \cdot p_2) \}$, where $p_4$ is the
momentum of the recoiling parton $a_4$ in the elastic-scattering subprocess
of Eq.~(\ref{elpro}). Using this expression for $u(q)$ in the BCMN resummed 
formulae, we directly obtain the
resummed expression of $\Sigma_{N}^{\,{\rm res}}$. Owing to their generality, 
the resummed formulae of Ref.~\cite{Bonciani:2003nt} are limited to the explicit
treatment of resummation to NLL accuracy. However, the specific kinematical
features of single-particle production near threshold \cite{Laenen:1998qw, 
Catani:1998tm, Bonciani:2003nt}
allow us to formally extend the validity of the resummation formulae obtained
from Ref.~\cite{Bonciani:2003nt} to arbitrary logarithmic accuracy.
The final result for the resummed radiative function $\Sigma_{N}^{\,{\rm res}}$
is presented below.

\begin{figure}[htb]
\begin{center}
\includegraphics[scale=0.6]{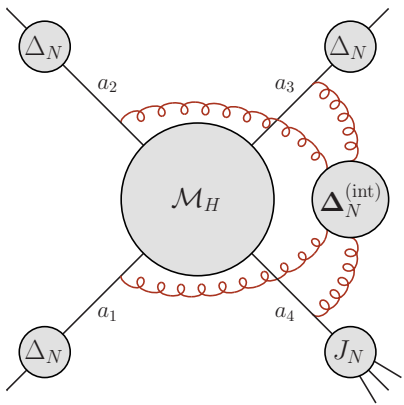}
\caption{
Pictorial representation of the all-order resummation formula in Eq.~(\ref{resu}).}
\label{fig1}
\end{center}
\end{figure}

The all-order resummation formula has a factorized structure
(Fig.~\ref{fig1}),
and it reads
\begin{equation}
\label{resu}
\Sigma_{a_1a_2\to a_3 a_4,N}^{\,{\rm res}}(r;p_T^2,\mu_F,\mu_f)
=\Bigg[\prod_{i=1,2,3}\Delta_{a_i,N_i}(Q_i^2;\mu^2_{Fi})\Bigg]\, J_{a_4,N_4}(Q_4^2)\,
\f{\langle {\cal M}_H| {\bom \Delta}^{\rm (int)}_N(r;p_T^2)|{\cal
M}_H\rangle}{|{\cal M}^{(0)}|^2}\, ,
\end{equation}
where ${\cal M}_H$ depends on the flavour indeces $a_i$ $(i=1,\dots,4)$, on the
kinematical variables $r$ and $p_T^2$, and on the factorization scales $\mu_F$
and $\mu_f$ (the Born-level scattering amplitude ${\cal M}^{(0)}$ depends on
$a_i$ and $r$). Each factor in the right-hand side of Eq.~(\ref{resu}) is
separately renormalization group invariant (i.e., it is independent of $\mu_R$
if it is evaluated to all orders in $\as(\mu_R^2)$).

The three radiative factors $\Delta_{a_i,N} \;(i=1,2,3)$
in the right-hand side of Eq.~(\ref{resu}) embody soft-gluon radiation from the
triggered partons $a_1, a_2$ and $a_3$ of the partonic process in 
Eq.~(\ref{partpro}). The $N$-moment factor $\Delta_{a,N}$ depends on the
flavour of the radiating parton $a$, on the partonic hard scale $Q^2$, and on
the factorization scale of the corresponding parton density or fragmentation
function in the hadronic cross section. We have
\begin{equation}
\label{delta}
\Delta_{a,N}(Q^2;\mu^2)= \exp \left\{ \int_0^1 dz\;\f{z^{N-1}-1}{1-z}\int_{\mu^2}^{(1-z)^2
Q^2} \f{dq^2}{q^2} \;A_a(\as(q^2)) \right\}\;\;,
\end{equation}
where $A_a(\as)$ is a perturbative function,
\begin{equation}
\label{Afun}
A_a(\as)=\sum_{n=1}^\infty \left(\f{\as}{\pi}\right)^n
A_a^{(n)}~~,
\end{equation}
whose lower-order coefficients are \cite{Catani:1989ne, Cacciari:2001cw}
\begin{equation}
\label{acoef}
A_a^{(1)}=C_a~,~~~~~A_a^{(2)}=\f{1}{2}C_a\,K~,~~~~~K=
C_A\left(\f{67}{18}-\f{\pi^2}{6}\right)-\f{5}{9}\,n_F~~,
\end{equation}
and the third-order coefficient $A_a^{(3)}$ is also known \cite{Moch:2004pa}
($A_a^{(3)}$ is the coefficient of the soft part of the 
DGLAP splitting function $P_{aa}(z,\as)$ at ${\cal O}(\as^3)$).

The jet function $J_{a_4,N_4}$ in Eq.~(\ref{resu}) 
includes soft and 
collinear radiation from the parton $a_4$ that recoils against the observed 
parton $a_3$ in the tree-level (or, more generally, elastic scattering) process 
$a_1a_2\to a_3 a_4$. The jet function  $J_{a,N}$, which depends 
on the flavour of the radiating parton $a$ and on the partonic hard scale $Q^2$,
has the following all-order form:
\begin{equation}
\label{jetf}
J_{a,N}(Q^2)=\exp \left\{ 
\int_0^1 dz\;\f{z^{N-1}-1}{1-z}\;\Big[\int_{(1-z)^2Q^2}^{(1-z)Q^2} \f{dq^2}{q^2} A_a(\as(q^2))+\f{1}{2}B_a\left(\as\left((1-z)Q^2\right)\right)\Big]
\right\} \, ,
\end{equation}
where $A_a(\as)$ is the same perturbative function as in Eqs.~(\ref{delta})
and (\ref{Afun}), and the perturbative function $B_a(\as)$ is
\begin{equation}
\label{Bfun}
B_a(\as)=\sum_{n=1}^\infty \left(\f{\as}{\pi}\right)^n
B_a^{(n)} \;\;,
\end{equation}
with the first-order coefficient 
\cite{Sterman:1986aj, Catani:1989ne, Catani:1990rp}
\begin{equation}
\label{bcoef}
B_a^{(1)}=-\gamma_a \;\;,
\end{equation}
where $\gamma_a$ is the same flavour coefficient as in Eq.~(\ref{gamcoef}).


The values of $N_i$ and $Q_i^2$ $(i=1,\dots,4)$ in the argument of the radiative
factors  $\Delta$ and $J$ in Eq.~(\ref{resu}) depend on $r, p_T^2$ and on the
moment index $N$ of $\Sigma_{N}^{\,{\rm res}}$. The specification of this
dependence involves some degree of arbitrariness
(see Ref.~\cite{Bonciani:2003nt}) that is compensated by a corresponding
dependence in the terms ${\bom \Delta}_N^{\rm (int)}$ and ${\cal M}_H$. We use
the Mellin moment values
\beq
\label{melni}
N_1= 
N \,\f{r}{1+r} \;, \quad
N_2= 
N \,\f{1}{1+r} \;, \quad
N_3= N \;, \quad
N_4 = 
N \,\f{r}{(1+r)^2} \;\;,
\eeq
and the common scale
\beq
\label{scaleqi}
Q^2_i = p_T^2,  \quad \quad i=1,2,3,4 \;\;,
\eeq
which unambiguously specify the expressions of 
${\bom \Delta}_N^{\rm (int)}$ and ${\cal M}_H$ that are presented below.

The radiative factors $\Delta_{a_i,N_i}$ and $J_{a_4,N_4}$ are $c$-number
functions. The term $|{\cal M}_H\rangle$ is a colour space vector (analogously
to the scattering amplitude $|{\cal M}\rangle$ in Eq.~(\ref{ampl}))
and ${\bom \Delta}^{\rm (int)}$ is a colour space operator (matrix) that acts on
$|{\cal M}_H\rangle$. Therefore, the last factor in the right-hand side of 
Eq.~(\ref{resu}) has a factorized structure in colour space, and it includes all
the colour correlation effects.

The colour-space radiative factor ${\bom \Delta}_N^{\rm (int)}$ embodies all the
quantum-interference effects that are produced by soft-gluon radiation at large 
angles with respect to the direction of the momenta $p_i$ $(i=1,\dots,4)$
of the partons in the $2 \to 2$ hard scattering.
Its explicit expression is \cite{Bonciani:2003nt}
\begin{equation}
\label{deltaint}
{\bom \Delta}^{\rm (int)}_N(r;p_T^2) ={\bom V}_N^{\,\dagger}(r;p_T^2)
\;{\bom V}_N(r;p_T^2) \;\;,
\end{equation}
where 
\begin{equation}
\label{bomV}
{\bom V}_N(r;p_T^2) = P_z\, \exp\left\{\int_0^1 dz\;\f{z^{N-1}-1}{1-z}
\,{\bom \Gamma}\left(\as\left((1-z)^2 p_T^2\right);r\right)\right\} \;\;.
\end{equation}
The soft-gluon anomalous dimension ${\bom \Gamma}(\as;r)$ is a colour space
matrix, and the operator $P_z$ 
denotes $z$-ordering in the expansion of the exponential matrix.
Note that the explicit expression of ${\bom \Gamma}$ can be changed by adding an
imaginary $c$-number contribution. This added term in  ${\bom \Gamma}$
produces an overall ($c$-number) phase factor in ${\bom V}_N$, and its effect is
cancelled by ${\bom V}_N^{\,\dagger}$ in the expression (\ref{deltaint}) of
${\bom \Delta}^{\rm (int)}_N$. 
Therefore, any imaginary $c$-number contribution to ${\bom \Gamma}$ is harmless,
since it has no effect on ${\bom \Delta}_N^{\rm (int)}$.
The anomalous-dimension matrix ${\bom \Gamma}(\as;r)$ has the perturbative expansion 
\begin{equation}
\label{bomGamma}
{\bom \Gamma}(\as;r)=\f{\as}{\pi}\,{\bom \Gamma}^{(1)}(r)
+\sum_{n=2}^\infty \left(\f{\as}{\pi}\right)^n
{\bom \Gamma}^{(n)}(r) \;\;,
\end{equation}
and the explicit expression of the first-order term is
\beeq
\label{gamma1}
{\bom \Gamma}^{(1)}(r) 
&=&  {\bf T}_t^2 \;\ln(1+r) + 
{\bf T}_u^2 \;\ln\f{1+r}{r}
+ i \pi \; {\bf T}_s^2 \;\\
\label{gamma1tu}
&=& {\bf T}_t^2 \;\Bigl( \ln(1+r) - i \pi \Bigr) 
+ {\bf T}_u^2 \;\Bigl( \ln\f{1+r}{r} - i \pi \Bigr)
+ i \pi \;\sum_{i=1}^4 C_{a_i} \;.
\eeeq
Note that ${\bom \Gamma}^{(1)}$ includes colour correlations, which we have
explicitly expressed in terms of the colour correlation operators in
Eqs.~(\ref{corcharge}) and (\ref{sumcorr}).
Note also that ${\bom \Gamma}^{(1)}(r)$ and, more generally, 
${\bom \Gamma}(\as;r)$ depend on the kinematical (angular) variable $r$,
at variance with the kernels $A_a(\as)$ and $B_a(\as)$
(which are independent of the kinematics) of $\Delta_{a_i,N_i}$ and 
$J_{a_4,N_4}$ in Eqs.~(\ref{delta}) and (\ref{jetf}).

The amplitude $|{\cal M}_H\rangle$ depends on the flavour, colour and kinematical
variables of the elastic scattering process $a_1a_2 \to a_3a_4$ in 
Eq.~(\ref{elpro}),
and it is independent of the Mellin moment $N$ (in practice, 
$|{\cal M}_H\rangle$ embodies the residual terms of  
$\Sigma_{N}^{\,{\rm res}}$ that are constant, i.e. of ${\cal O}(1)$, and not
logarithmically enhanced in the limit $N \to \infty$).
The colour space amplitude $|{\cal M}_H\rangle$ has an all-order 
perturbative structure that is analogous to the structure of the scattering
amplitude $|{\cal M}\rangle$ in Eq.~(\ref{ampl}).
We write
\begin{equation}
|{\cal M}_H\rangle= 
\as(\mu_R^2)\left[|{\cal M}^{(0)}\rangle + \f{\as(\mu_R^2)}{2\pi}\;|{\cal M}^{(1)}_H(\mu_R)\rangle
+\sum_{n=2}^\infty\left(\f{\as(\mu_R^2)}{2\pi}\right)^n|{\cal
M}^{(n)}_H(\mu_R)\rangle\right] \;\;,
\end{equation}
where we have omitted the explicit reference to the parton indices
$a_1a_2a_3a_4$. At the lowest order, ${\cal M}_H$ exactly coincides with the
Born-level scattering amplitude ${\cal M}^{(0)}$ 
($\,|{\cal M}^{(0)}_H\rangle=|{\cal M}^{(0)}\rangle$).
The analogy between ${\cal M}_H$ and ${\cal M}$ persists at higher orders, since
${\cal M}_H$ also refers to the elastic scattering $a_1a_2 \to a_3a_4$ and it can
be regarded as the `hard' (i.e., IR finite) component of the virtual
contributions to the renormalized scattering amplitude  ${\cal M}$.
The amplitude $|{\cal M}_H\rangle$ is obtained from $|{\cal M}\rangle$ by
removing its IR divergences and a definite amount of IR finite terms.
The (IR divergent and finite) terms that are removed from $|{\cal M}\rangle$ originate from the (soft) 
real emission contributions to the cross section and, therefore, these terms
and $|{\cal M}_H\rangle$ specifically depend on the one-parton inclusive cross
section (i.e., $|{\cal M}_H\rangle$ is an observable-dependent quantity).

The first-order term ${\cal M}^{(1)}_H$ of ${\cal M}_H$ can be obtained from the
results of our NLO calculation of the partonic cross section near threshold.
We consider the NLO result in Eqs.~(\ref{nlofactform}) and (\ref{C1ris}),
and we use Eq.~(\ref{plusnlo}) to explicitly perform the change of variables
$\{ s, v, w \} \to \{ \xw, r, p_T^2 \}$.
Then we can compute its $N$-moments with respect to $\xw$ and, in the limit
$N \to \infty$, we compare this result with the perturbative expansion of
the resummation formula in Eq.~(\ref{resu}) up to relative 
${\cal O}(\as)$. From the comparison we cross-check that the 
structure and the coefficients of the logarithmic terms ($\ln^2N$ and $\ln N$)
do agree, and we can extract $|{\cal M}^{(1)}_H\rangle$ in explicit form.
Note that the colour-space factorization form of our NLO result 
in Eqs.~(\ref{nlofactform}) and (\ref{C1ris}) is essential to obtain
the {\em amplitude} $|{\cal M}^{(1)}_H\rangle$.
The ${\cal O}(\as)$ expansion of the resummation formula (\ref{resu})
can also be compared with the analytic NLO result of Ref.~\cite{Aversa:1988vb}.
However, the latter contains only the `colour-summed' function
${\cal C}^{(1)}$ of Eq.~(\ref{C1}) and, therefore, from the comparison we can
only extract the colour-summed interference 
`$\langle {\cal M}^{(0)} | {\cal M}^{(1)}_H\rangle + {\rm c.c.}$'.
The knowledge of the colourful amplitude  $|{\cal M}^{(1)}_H\rangle$
(rather than `$\langle {\cal M}^{(0)} | {\cal M}^{(1)}_H\rangle + {\rm c.c.}$')
is important for QCD predictions beyond the NLO
(see also Eq.~(\ref{2loopinter}) and related comments).

Our first-order result for the hard-virtual amplitude ${\cal M}_H$ is
\begin{equation}
\label{MH1}
|{\cal M}^{(1)}_H\rangle=|{\cal M}^{(1)}\rangle-{\bom I}^{(1)}_H \;|{\cal
M}^{(0)}\rangle\; ,
\end{equation}
where $|{\cal M}^{(1)}\rangle$ is the full one-loop scattering amplitude in
Eqs.~(\ref{ampl}) and (\ref{m1fin}), and the colour space operator 
${\bom I}^{(1)}_H$ has the following explicit form:
\begin{align}
\label{IH1}
{\bom I}^{(1)}_H&={\bom I}^{(1)}_{\rm sing}+
\f{\pi^2}{4}\left(\T_1^2+\T_2^2+\T_3^2+\f{4}{3}\T_4^2\right)+\f{1}{2}\sum_{i=1}^3\gamma_{a_i}\ln\frac{\mu_{Fi}^2}{p_T^2}\nn\\
&-\f{1}{2}\ln (1+r)\ln\f{1+r}{r} \left(\T_1^2+\T_2^2-3\T_3^2+\T_4^2\right)\nn\\
&-\T_t^2 \left(\f{\pi^2}{2}+\f{1}{2}\ln^2(1+r)+\ln(1+r)\ln\f{1+r}{r}\right)\nn\\
&-\T_u^2
\left(\f{\pi^2}{2}+\f{1}{2}\ln^2\f{1+r}{r}+\ln(1+r)\ln\f{1+r}{r}\right)-\f{1}{2}K_{a_4}
\\
\label{I1fin}
& \equiv {\bom I}^{(1)}_{\rm sing} + {\bom I}^{(1)\, {\rm fin}}_H
\;\; ,
\end{align}
where the operator ${\bom I}^{(1)}_{\rm sing}$ is defined in Eq.~(\ref{I1sing}) and the flavour coefficients
$K_a$ are given in Eq.~(\ref{Kappa}).
Note that Eq.~(\ref{MH1}) can be rewritten in terms of 
${\cal M}^{(1)\,{\rm fin}}$ in Eq.~(\ref{m1fin}) and of the IR-finite part 
${\bom I}^{(1) \,{\rm fin}}_H$ of ${\bom I}^{(1)}_H$ (as defined by 
Eq.~(\ref{I1fin})). We have
\begin{equation}
\label{MH1fin}
|{\cal M}^{(1)}_H\rangle=|{\cal M}^{(1)\,{\rm fin}}\rangle 
-  {\bom I}^{(1)\, {\rm fin}}_H\;|{\cal M}^{(0)}\rangle \; ,
\end{equation}
which explicitly shows that the one-loop hard-virtual amplitude 
${\cal M}_H^{(1)}$ is evidently IR finite.

We note that our all-order resummed results for $\Sigma_{N}^{\,{\rm res}}$
preserve the angular symmetry (i.e. the symmetry under the exchange 
$t \leftrightarrow u$) of the complete (i.e. without the near-threshold
approximation in Eq.~(\ref{Nsigmares})) partonic cross section. This
symmetry corresponds to the exchange 
$\{ r, {\bom T_1}, \gamma_{a_1} \} \leftrightarrow
\{ 1/r, {\bom T_2}, \gamma_{a_2} \}$, and it is manifest in our resummed formulae
(see Eqs.~(\ref{resu}), (\ref{delta}), (\ref{jetf}), (\ref{melni}),
(\ref{gamma1}), (\ref{IH1})).

The expressions (\ref{delta}), (\ref{jetf}) and (\ref{bomV}) of the radiative
factors $\Delta, J$ and ${\bom V}$ involve the integrations of $\as(k^2)$ over
the scale $k^2$ that (roughly) corresponds to the square of the relative
transverse momentum of each primary parton (which radiates subsequent final-state
partons) that is emitted from the Born-level process $a_1a_2\to a_3a_4$. This
transverse-momentum scale is then related 
(through the integration over $z$) to two characteristic scales of the
`measured' single-particle inclusive cross section, namely the `soft' scale
$(1-z)^2p_T^2$ and the `collinear' scale $(1-z)p_T^2$. The soft and collinear
scales are related to the total energy and recoiling invariant mass that
accompany the underlying Born-level process. The factorized structure of 
Eq.~(\ref{resu}) follows \cite{Bonciani:2003nt} from general features of soft and
collinear QCD radiation and, therefore, it has 
analogies with the structure of factorization formulae that are derived for 
kinematically-related processes
\cite{Becher:2009th, Becher:2011fc} by using SCET methods.

The expression (\ref{gamma1}) (or (\ref{gamma1tu}))
of the soft-gluon anomalous dimension ${\bom \Gamma}^{(1)}(r)$ is particularly
simple, though it still depends on two colour correlation operators (as recalled
in Sect.~\ref{sec:nlo}, this is the maximal number of linearly-independent colour
correlations in the case of $2 \to 2$ scattering).
The dependence on colour correlations simplifies in two kinematical
configurations that correspond to parton production at very small or very large
rapidities (in the centre--of--mass frame of the $2 \to 2$ partonic reaction). 
In the limit $r \to 1$, which corresponds (see Eq.~(\ref{kineta}))
to $\eta \simeq 0$ or $\theta^*_{13} \simeq 90^o$, 
Eq.~(\ref{gamma1}) gives
\beq
\label{gamma190}
{\bom \Gamma}^{(1)}(r=1) =
{\bf T}_s^2 \,\bigl( - \ln 2 +  i \pi \bigr) + \sum_{i=1}^4 C_{a_i} \,\ln 2 
\;\;,
\eeq
where we have used Eq.~(\ref{sumcorr}).
We note that the expression in Eq.~(\ref{gamma190}) depends on a {\em single}
colour correlation operator, which is simply 
${\bf T}_s^2= ({\bf T}_1 + {\bf T}_2)^2$, the square of the colour charge in the
$s$-channel.
In the limit $r \to \infty$, which corresponds to very forward production
of the parton $a_3$ ($\eta \to +\infty$,  or $\theta^*_{13} \to 0$), 
Eq.~(\ref{gamma1tu}) gives
\beq
\label{gamma1for}
{\bom \Gamma}^{(1)}(r) \simeq {\bf T}_t^2 \;\ln r \,=\,
{\bf T}_t^2 \;2 \eta \, 
\simeq {\bf T}_t^2 \;\ln \left( \f{2}{\theta^*_{13}} \right)^2
\;\;, \quad \quad ( r \gg 1 ) \;,
\eeq
and we see that ${\bom \Gamma}^{(1)}$ {\em only} depends on 
${\bf T}_t^2=({\bf T}_1 + {\bf T}_3)^2$, which is the square of the colour 
charge exchanged in the $t$-channel.
An analogous result is obtained in the case of very backward production of the
parton $a_3$ by performing the limit 
$r \to 0$ ($\eta \to -\infty$,  or $\theta^*_{23} \to 0$):
\beq
\label{gamma1back}
{\bom \Gamma}^{(1)}(r) \simeq {\bf T}_u^2 \;\ln \f{1}{r} \,= \,
{\bf T}_u^2 \;\left( - 2 \eta \right) \, 
\simeq {\bf T}_u^2 \;\ln \left( \f{2}{\theta^*_{23}} \right)^2
\;\;, \quad \quad ( r \ll 1 ) \;.
\eeq

The fact that a {\em single} colour correlation operator survives in the limits
of Eqs.~(\ref{gamma190}), (\ref{gamma1for}) and (\ref{gamma1back}) is not
accidental, and it is a general feature of $2 \to 2$ parton scattering. The
result in Eq.~(\ref{gamma1}) refers to a specific observable, namely, the
one-particle inclusive cross section. The {\em general} form of the soft-gluon
anomalous dimension at ${\cal O}(\as)$ for $2 \to 2$ parton scattering
can be written as \cite{Bonciani:2003nt, Dokshitzer:2005ig}
\beq
\label{gamma1gen}
{\bom \Gamma}^{(1)\,{\rm obs.}}(2 \to 2) 
=  {\bf T}_t^2 \;\ln \left(\f{s}{-t}\right) + 
{\bf T}_u^2 \;\ln \left(\f{s}{-u}\right)
+ i \pi \; {\bf T}_s^2 + \sum_{i=1}^4 C_{a_i} \,g_i^{{\rm obs.}}(s,t,u)\;\;,
\eeq
where $\{ s, t, u \}$ are the Mandelstam variables  of the $2 \to 2$ elastic
process. This expression is valid for a generic (`global') observable that is
dominated by soft-gluon radiation in $2 \to 2$ hard scattering
(the expression (\ref{gamma1gen}) is directly obtained by specifying the
$m$-parton expressions in Eqs.~(21) and (27) 
of Ref.~\cite{Bonciani:2003nt} to the case of $m=4$ hard partons).
The dependence of ${\bom \Gamma}^{(1)\,{\rm obs.}}$ on the observable is entirely
given by the functions $g_i^{{\rm obs.}}$, and it produces a $c$-number
contribution
(it is proportional to the Casimir coefficients $C_{a_i}$). The dependence of 
${\bom \Gamma}^{(1)\,{\rm obs.}}$ on the colour correlation operators is instead
universal (i.e., independent of the specific observable). Setting $-t=-u=s/2$
in Eq.~(\ref{gamma1gen}), ${\bf T}_s^2$ is the sole correlation operator that
appears in ${\bom \Gamma}^{(1)\,{\rm obs.}}$. Considering the limit $t \to 0$
($u \to 0$) of Eq.~(\ref{gamma1gen}), ${\bf T}_t^2$ (${\bf T}_u^2$)
is the sole correlation operator that
appears in ${\bom \Gamma}^{(1)\,{\rm obs.}}$. These are exactly the colour 
correlation operators that are singled out by the corresponding limits in
Eqs.~(\ref{gamma190}), (\ref{gamma1for}) and (\ref{gamma1back}).

We also note that, in the large rapidity limits of 
Eqs.~(\ref{gamma1for}) and (\ref{gamma1back}), the expression of 
${\bom \Gamma}^{(1)}(r)$ is {\em especially} simple: it is simply proportional to
${\bf T}_t^2$ or ${\bf T}_u^2$ (as predicted by Eq.~(\ref{gamma1gen})) with {\em
no} additional $c$-number contributions. This simplicity is due to the scale
choice $Q_i^2=p_T^2$ in Eq.~(\ref{scaleqi}), and it has a direct interpretation
as colour coherence phenomenon. The illustration of colour coherence is
particularly simple for the production at large rapidities, since we can neglect
effects of ${\cal O}(1/\eta)$.
In the case of very forward production ($\theta^*_{13} \to 0$), each of the four
hard partons $a_i$ ($i=1,\dots,4$) radiates soft partons (interjet radiation)
as an independent emitter
(with intensity proportional to its colour charge ${\bf T}_i^2=C_{a_i}$
as in Eqs.~(\ref{delta}) and (\ref{jetf}))
 {\em inside} a small angular region of size 
$\theta^*_{13}=\theta^*_{24}$ around the direction of its momentum. Soft-parton
radiation at larger angles (intrajet radiation) feels the coherent action of the
`forward emitter' (the pair of partons $a_1$ and $a_3$, which are seen as two
exactly collinear partons, i.e. as a single parton,
by radiation at wide angles) and of the `backward emitter'
(the pair of partons $a_2$ and $a_4$). The forward and backward emitters radiate
(independently) with intensity proportional to their colour charge 
${\bf T}_t^2=({\bf T}_1 + {\bf T}_3)^2=({\bf T}_2 + {\bf T}_4)^2$
over the wide-angle region, which occupies the large rapidity interval of size
$2 \eta$: this leads to the radiation probability ${\bf T}_t^2 \; 2 \eta$
(see Eq.~(\ref{gamma1for})). This colour coherence picture corresponds to the
factorization structure of Eqs.~(\ref{resu}) in terms of the corresponding
radiative factors. The absence of terms proportional to ${\bf T}_i^2=C_{a_i}$ in 
Eq.~(\ref{gamma1for}) implies that ${\bf \Delta}^{({\rm int})}$ exactly
originates from intrajet radiation, while interjet radiation is exactly included 
in each of the four radiative factors $\Delta_{a_i}(Q_i^2)$ and $J_{a_4}(Q_4^2)$.
Indeed, in the limit $\theta^*_{13} \to 0$ ($p_T \simeq p_3^0 \theta^*_{13}$),
the transverse-momentum scales $(1-z)^2 Q_i^2 =(1-z)^2 p_T^2$ 
in Eq.~(\ref{scaleqi}) precisely correspond to radiation from $p_i$ up to a
maximum angle $\theta_{\rm max}\simeq\theta^*_{13} =  \theta^*_{24}$.

The factorized structure of 
$\langle {\cal M}_H| {\bom \Delta}^{\rm (int)}_N |{\cal M}_H\rangle$
in colour space entails colour {\em interferences between} 
${\bom \Delta}^{\rm (int)}_N$ and $|{\cal M}_H\rangle$.
The colour interference effects start to contribute at ${\cal O}(\as^2)$.
The dominant logarithmic terms in ${\bom \Delta}^{\rm (int)}_N$ are of 
${\cal  O}(\as^n \ln^n N)$, and we can consider the following approximation:
\begin{equation}
\label{nointapp}
\langle {\cal M}_H| {\bom \Delta}^{\rm (int)}_N|{\cal M}_H\rangle\to 
\langle {\cal M}^{(0)}| {\bom \Delta}^{\rm (int)}_N|{\cal M}^{(0)}\rangle
\times \f{|{\cal M}_H|^2}{|{\cal M}^{(0)}|^2}\, 
+ {\cal  O}\left(\as(\as\ln N)^n\right) \;\;,
\end{equation}
which shows that the colour interference effects can be neglected up to
${\cal O}((\as\ln N)^n)$.
Starting from ${\cal  O}\left(\as(\as\ln N)^n\right)$, 
the colour interference effects
are relevant. In particular, 
$\langle {\cal M}_H| {\bom \Delta}^{\rm (int)}_N |{\cal M}_H\rangle$
leads to the second-order contribution
\beq
\label{2loopinter}
-\f{1}{2} \left(\f{\as}{\pi}\right)^2 \,\ln N
\left\{ \,2 \,\langle {\cal M}^{(0)}| 
\left( {\bom \Gamma}^{(2)} + {\bom \Gamma}^{(2) \,\dagger} \right) 
\,|{\cal M}^{(0)}\rangle + \left(
\langle {\cal M}^{(0)}| 
\left( {\bom \Gamma}^{(1)} + {\bom \Gamma}^{(1) \,\dagger} \right) 
|{\cal M}_H^{(1)}\rangle + {\rm c.c.}
\right)
\right\}
\eeq
that is incorrectly approximated by neglecting colour interferences as in the
right-hand side of Eq.~(\ref{nointapp}). Using the approximation in 
Eq.~(\ref{nointapp}), the last term in the curly bracket of 
Eq.~(\ref{2loopinter}) would be replaced by 
$\langle {\cal M}^{(0)}| 
\left( {\bom \Gamma}^{(1)} + {\bom \Gamma}^{(1) \,\dagger}\right)
|{\cal M}^{(0)}\rangle \left( \langle {\cal M}^{(0)}|{\cal M}_H^{(1)}\rangle
+ {\rm c.c.}\right)/|{\cal M}^{(0)}|^2$. The expression in Eq.~(\ref{2loopinter})
explicitly shows that the second-order anomalous dimension 
${\bom \Gamma}^{(2)}$ contributes at the same level of logarithmic accuracy as
the colour interference between ${\bom \Gamma}^{(1)}$ and 
$|{\cal M}_H^{(1)}\rangle$.

The all-order structure of Eqs.~(\ref{resu}), (\ref{delta}), (\ref{jetf}) 
and (\ref{bomV}) leads to the resummation of the $\ln N$ terms in {\em
exponentiated} form. In Eq.~(\ref{bomV}), exponentiation has a formal meaning,
since it refers to the formal exponentiation of matrices. However, the
anomalous dimension matrix ${\bom \Gamma}(\as;r)$ can be {\em diagonalized}
\cite{Kidonakis:1998nf, Dokshitzer:2005ig} in colour space.
After diagonalization, the resummed radiative function 
$\Sigma_{N}^{\,{\rm res}}$ of Eq.~(\ref{resu}) can be written in the customary
(see, e.g., Refs.~\cite{Catani:1998tm, deFlorian:2005yj})
exponential form
\beeq
\label{exponfin}
\Sigma_{a_1a_2\to a_3 a_4,N}^{\,{\rm res}}(r;p_T^2,\mu_F,\mu_f) =&&
\!\!\!\!\!\!\!\!\!\!\!\!\!\!
\sum_{I} \;
{\widetilde C}_{I,a_1a_2a_3 a_4}(\as(p_T^2),r;p_T^2,\mu_F,\mu_f) \nn \\
&\times&\!\! \exp \Bigl\{ 
{\cal G}_{I,a_1a_2a_3 a_4}(\as(p_T^2),\ln N, r;p_T^2,\mu_F,\mu_f)
 \Bigr\} 
+ {\cal O}\left(\f{1}{N}\right) \,,
\eeeq
where the index $I$ labels the colour space eigenstates 
$|I(\as;r)\rangle$ of ${\bom \Gamma}(\as;r)$, and 
${\widetilde C}$ and ${\cal G}$ are functions (they are not colour matrices).
These functions are renormalization group invariant, and
their dependence on $\mu_R$ arises by writing $\as(p_T^2)$ as a function of
$\as(\mu_R^2)$ and $\ln (p_T^2/\mu_R^2)$ (as in customary perturbative
calculations).

The exponent function ${\cal G}_I$ includes all the $\ln N$ terms, and it can
consistently be expanded in LL terms of ${\cal  O}(\as^n \ln^{n+1} N)$,
NLL terms of ${\cal  O}(\as^n \ln^{n} N)$,
NNLL terms of ${\cal  O}(\as(\as \ln N)^n)$, and so forth. 
The function ${\widetilde C}_I$ does not depend on $N$, since it includes all the
terms that are constant (i.e., of ${\cal  O}(1)$) in the limit $N \to \infty$.
The LL terms of ${\cal G}_I$ (they are actually independent of $I$) are
controlled by the perturbative coefficient $A_a^{(1)}$ in Eq.~(\ref{acoef}).
The NLL terms of ${\cal G}_I$ are then fully determined by 
$A_a^{(2)}$ (see Eq.~(\ref{acoef})), $B_a^{(1)}$ (see Eq.~(\ref{bcoef}))
and ${\bom \Gamma}^{(1)}(r)$ in Eq.~(\ref{gamma1}) (or, more precisely, the
eigenvalues $\Gamma^{(1)}_I(r)$ of ${\bom \Gamma}^{(1)}$). 
The Born-level contribution to the function ${\widetilde C}_I$ depends on
$|\langle I | {\cal M}^{(0)} \rangle|^2$. The first-order term
${\widetilde C}_I^{(1)}$ of the function ${\widetilde C}_I$ depends on
`$\langle {\cal M}^{(0)}| I \rangle \langle \,I |{\cal M}_H^{(1)} \rangle + {\rm
c.c.}$', and 
this colour interference (between $|{\cal M}^{(0)}\rangle \,, |I\rangle$ and 
$|{\cal M}_H^{(1)} \rangle$)
is computable from the explicit expression of 
$|{\cal M}_H^{(1)} \rangle$ 
(see Eqs.~(\ref{MH1}) and (\ref{IH1})). Since we know 
${\bom \Gamma}^{(1)}$ and $|{\cal M}_H^{(1)} \rangle$, the colour interference between
these two terms (see Eq.~(\ref{2loopinter})) is known
(in the colour-diagonalized expression (\ref{exponfin}), the interference is
taken into account by the correlated dependence on $I$ between 
${\widetilde C}_I^{(1)}$ and $\Gamma^{(1)}_I$ in the exponent ${\cal G}_I$).
Therefore, the complete explicit
determination of the NNLL terms in ${\cal G}_I$ still requires the coefficient 
$A_a^{(3)}$ in Eq.~(\ref{Afun}) (this coefficient is known \cite{Moch:2004pa}),
the coefficient $B_a^{(2)}$ in Eq.~(\ref{Bfun}) and the second-order anomalous
dimension ${\bom \Gamma}^{(2)}(r)$ in Eq.~(\ref{bomGamma}).
The bulk of the contributions to ${\bom \Gamma}^{(2)}(r)$ is expected
\cite{Catani:1990rr, Catani:1998bh, Bonciani:2003nt, Aybat:2006mz}
to be proportional to ${\bom \Gamma}^{(1)}(r)$ and obtained by inserting the
simple rescaling
(the coefficient $K$ is given in Eq.~(\ref{acoef})) 
\beq
\label{cmw}
\as \to \as \left[ 1 + \f{\as}{\pi} \,\f{1}{2}\, K \right]
\eeq
in the expression of 
${\bom \Gamma}$ at ${\cal  O}(\as)$. The coefficient $B_a^{(2)}$ could be
extracted from NNLL computations of related processes, such as DIS 
\cite{Vogt:2000ci, Moch:2005ba, Soar:2009yh}
and direct-photon production \cite{Becher:2009th}.

We have previously noticed that the explicit contributions of the radiative
factors to the factorization formula (\ref{resu}) depend on the choice of the
scales $Q_i^2$. The specific choice in Eq.~(\ref{scaleqi}) is mostly a matter of
convenience (in the case of direct-photon production, for instance, the 
scales $Q_i^2$ can be chosen in such a way that the first-order
term $\Gamma^{(1)}$ vanishes \cite{Catani:1998tm}) and, possibly, of closer
correspondence with colour coherence features. We remark that, combining the
radiative factors, the complete result for 
$\Sigma_{N}^{\,{\rm res}}$ is fully independent of the scales $Q_i^2$. In
particular, the all-order expression in Eq.~(\ref{exponfin}) and the functions 
${\cal G}_I$ and ${\widetilde C}_I$ are fully independent of $Q_i^2$, and this
independence persists after the consistent truncation at arbitrary 
N$^k$LL accuracy (and/or at arbitrary orders in $\as$).

We also note that the exponentiated expressions 
(\ref{delta}), (\ref{jetf}) and (\ref{bomV}) of the radiative factors can be
rewritten in a different, though eventually equivalent, form. This alternative
form is obtained by the replacement
\beq
\int_0^1 dz\;\f{z^{N-1}-1}{1-z} \;\dots \to 
- \int_0^{1- N_0/N} \;\f{dz}{1-z}\;\dots
\eeq
where $N_0= e^{-\gamma_E}$ 
($\gamma_E = 0.5772\dots$ is the Euler number). The replacement is directly valid
up to NLL accuracy \cite{Catani:1989ne}, and it is also applicable to arbitrary
logarithmic accuracy (see Ref.~\cite{Catani:2003zt} for the related details),
provided the functions $B_a(\as)$ 
and ${\bom \Gamma}(\as;r)$ are correspondingly (and properly) redefined
starting from ${\cal O}(\as^2)$
(some non-logarithmic terms have to be reabsorbed in $M_H$, starting from $M_H^{(1)}$).
We remark that this alternative representation of
the radiative factors leads to the same (all-order) results for 
$\Sigma_{N}^{\,{\rm res}}$ and for the functions 
${\cal G}_I$ and ${\widetilde C}_I$ in Eq.~(\ref{exponfin})
(different representations can only lead to differences of 
${\cal O}(1/N)$).

The results that we have presented in Sect.~\ref{sec:nlo}
and in this section for the unpolarized scattering reaction in 
Eq.~(\ref{hadpro})
{\em equally} apply to processes in which one or more of the three triggered
partons $a_1, a_2$ and $a_3$ (hadrons $h_1, h_2$ and $h_3$) are spin polarized.
The relation between the unpolarized and polarized cases is technically
straightforward within the process-independent formalism that we have used and
explicitly worked out. Since the NLO results of Sect.~\ref{sec:nlo}
are embodied in the ${\cal O}(\as)$ expansion of the resummed results,
we comment on polarized processes by simply referring to the results presented
in this section. In particular, we only remark the technical differences that
occur in the final results.
 
The unpolarized partonic cross section in Eq.~(\ref{Sigmadef})
is replaced by the spin-polarized cross section, and analogous replacement
applies to the factors ${\sigma}^{(0)}$ and $\Sigma$ in the right-hand side.
Obviously, the Born-level factor in Eq.~(\ref{sigborn1}) 
has to be computed by replacing the spin-averaged factor 
$|{\overline {{\cal M}^{(0)}}|^2}$
with the corresponding spin-dependent factor.
Note that the polarized cross section can acquire an explicit dependence on the
azimuthal angle $\phi$
of the two-dimensional transverse-momentum vector of the triggered 
parton/hadron
(for instance, this dependence occurs in the case of collisions of
transversely-polarized hadrons). The structure of the all-order resummation
formula (\ref{resu})
is unchanged in the polarized case. In particular, the soft-gluon radiative
factors $\Delta_{a_i,N_i}$ ($i=1,2,3$) and ${\bom \Delta}_N^{(\rm int)}$
in Eq.~(\ref{resu}) are exactly the same for both the unpolarized and polarized
cases. This is a consequence of spin-independence of soft-gluon radiation. The
dependence on the spin polarizations may enter into the resummation formula only
through the factors $J_{a_4}$, $|{\cal M}_H \rangle$ and 
$|{{\cal M}^{(0)}|^2}$ in Eq.~(\ref{resu}).
We shortly comment on this dependence.

Collinear radiation is possibly sensitive to spin and spin-polarizations. In the
resummation formula (\ref{resu}), collinear radiation is embodied in the jet
function $J_{a_4}$ and in ${\cal M}_H$ (e.g., through the collinear coefficient
$K_{a_4}$ in Eq.~(\ref{IH1})). These collinear contributions arise from the
small-mass recoiling jet $X$ in the inclusive process of Eq.~(\ref{partpro}).
Since the final-state partons in the system $X$ are inclusively summed
(including the sum over their spin polarizations), the ensuing collinear
contributions do not depend on the polarization of the triggered partons 
$a_1, a_2$ and $a_3$. Therefore, in the resummation formula (\ref{resu}),
the only source of spin-polarization dependence is in the hard-radiation
contributions embodied in ${\cal M}_H$ 
(and in $|{\cal M}^{(0)}|^2$).

Throughout the paper, using the bra-ket notation 
$\langle\, \dots | \dots \rangle$ we have denoted the sum over colours and
implicitly assumed a sum over the spin polarization states of the partons 
$a_i$ of the partonic subprocess $a_1 a_2 \to a_3 a_4$. In the case of polarized
scattering, we can simply release this implicit assumption, and the products
$\langle\, \dots | \dots \rangle$ are computed by using  
$|{\cal M}^{(0)}\rangle$ 
($|{\cal M}^{(0)}|^2= \langle {\cal M}^{(0)} | {\cal M}^{(0)}\rangle$)
and $|{\cal M}_H \rangle$ at fixed spin-polarization states of one or more of
the three partons $a_1, a_2$ and $a_3$ (according to the definite polarization
states of the scattering process of interest). The spin dependence of the
tree-level (${\cal M}^{(0)}$) and one-loop (${\cal M}^{(1)}$) amplitudes 
is known \cite{Bern:1991aq, Kunszt:1993sd}.
This directly determines  the spin dependence of the one-loop hard-virtual 
$|{\cal M}_H^{(1)} \rangle$ in Eq.~(\ref{MH1}),
and the ensuing spin dependence  of the resummation formula (\ref{resu}).
We simply note that the computation of radiative corrections for 
polarized cross sections involves customary
$d$-dimensional subtleties related to spin. We refer, for instance, to variants
for the $d$-dimensional treatment of the Dirac matrix $\gamma_5$ 
\cite{Vogelsang:1996im},
and to the use of variants of dimensional regularization 
\cite{Kunszt:1993sd, Catani:1996pk}
for treating gluon polarizations. This spin-related subtleties in the
computation of the one-loop amplitude ${\cal M}^{(1)}$ have to be treated 
in a fully consistent manner (or, consistently related 
\cite{Kunszt:1993sd, Vogelsang:1996im, Catani:1996pk})
to avoid any ensuing mismatch in the computation of ${\cal M}_H^{(1)}$
according to Eq.~(\ref{MH1}). We recall that our explicit expression
(\ref{IH1}) (and, in particular, the value of the 
coefficient $K_{a_4}$ in the corresponding
Eq.~(\ref{Kappa})) of the operator ${\bom I}_H^{(1)}$ refers to the use of the
customary CDR scheme.

Soft-gluon resummation at NLL accuracy for single-hadron inclusive production 
in collisions of longitudinally-polarized and transversely-polarized hadrons
has been performed in Ref.~\cite{de Florian:2007ty}. The resummation study of
Ref.~\cite{de Florian:2007ty} deals with the rapidity-integrated cross section.

The rapidity-integrated cross section is considered in the following
Sect.~\ref{sec:rapint}. Owing to the straightforward relation between our
resummed formulae for the unpolarized and polarized cases (as we have just
discussed), in Sect.~\ref{sec:rapint} we limit ourselves to explicitly 
referring to the unpolarized case.

\subsection{Cross section integrated over the rapidity}
\label{sec:rapint}

A soft-gluon resummation formula that is similar to Eq.~(\ref{resu})
can be written in the kinematically-simpler case 
\cite{Catani:1998tm, Sterman:2000pt, Bonciani:2003nt, deFlorian:2005yj} 
in which the single-particle cross section is integrated over
the rapidity of the observed hadron (parton).
This rapidity-integrated resummation formula can be obtained from 
Eq.~(\ref{resu}).

To show the relation between these resummation formulae,
we consider the hadronic cross section
$d\sigma_{h_3}/d^2\bP_{3T}$, which is obtained by integrating the differential
cross section in Eq.~(\ref{factorization}) over the rapidity of the observed
hadron $h_3(P_3)$. The form of the QCD factorization formula 
(\ref{factorization}) is unchanged, although the partonic cross section 
$p^0_3d{\hat \sigma}/{d^3\bp_3}$ is replaced by the corresponding partonic cross
section $d{\hat \sigma}/d^2\bp_{T}$, which is obtained by integration of 
$p^0_3d{\hat \sigma}/{d^3\bp_3}$ over the rapidity $\eta$ of the parton $a_3$
at fixed value of its transverse momentum $\bp_{T}=\bp_{3T}$.
By analogy with Eq.~(\ref{Sigmadef}) we define
\begin{equation}
\label{Sigmadefeta}
\f{d{\hat \sigma}_{a_1a_2\to a_3}}{d^2\bp_{T}}=\f{1}{(4 \pi s)^2}\,
{\widetilde \Sigma}_{a_1a_2\to a_3}(x_T;p_T^2, \mu_F, \mu_f) \;\;,
\end{equation}
where the function ${\widetilde \Sigma}$ is dimensionless, and the kinematical
variable $x_T$ is the customary scaling variable
\begin{equation}
\label{xtdef}
x_T=\f{2p_T}{\sqrt s} \;\;.
\end{equation}
We recall (see Eq.~(\ref{kineta}))
that the variables $\xw$ and $x_T$ are related through the rapidity
$\eta$ by the kinematical relation
\begin{equation}
\label{xtvseta}
x_\omega=x_T \cosh \eta \;\;.
\end{equation}
The dependence of the cross section on the Born-level amplitude
${\cal M}^{(0)}$ (we recall that $|{\cal M}^{(0)}|^2$ only depends on $r=e^{2\eta}$)
is included in ${\widetilde \Sigma}$, so that the perturbative QCD expansion of
${\widetilde \Sigma}$ has the following overall normalization:
\begin{equation}
\label{wspert}
{\widetilde \Sigma}_{a_1a_2\to a_3}(x_T;p_T^2, \mu_F, \mu_f)
= \as^2(\mu_R^2) \,{\widetilde \Sigma}_{a_1a_2\to a_3}^{(0)}(x_T) \Bigl[ 1 +
{\cal O}(\as(\mu_R^2)) \Bigr] \;\;,
\end{equation}
where
\begin{equation}
\label{bornetaint}
{\widetilde \Sigma}_{a_1a_2\to a_3}^{(0)}(x_T) = \int d\eta \;\;
|{\overline {{\cal M}_{a_1a_2a_3a_4}^{(0)}(r=e^{2\eta})}}|^2
\;\delta(1 - x_T \cosh \eta) \;\;.
\end{equation}

In the case of the $p_T$-dependent cross section of 
Eq.~(\ref{Sigmadefeta}), the region of partonic threshold corresponds to the
limit $x_T \to 1$. In this limit, the higher-order radiative corrections to
${\widetilde \Sigma}(x_T)$ are logarithmically enhanced: the contributions in the
square bracket of Eq.~(\ref{wspert}) include terms of the type 
$\as^n \ln^m (1-x_T)$ (with $m \leq 2n$) (these terms arise from the rapidity
integration of the plus-distributions of the variable $\xw$). The all-order
resummation of these terms is performed in Mellin space by introducing the $N$
moments ${\widetilde \Sigma}_N$ of ${\widetilde \Sigma}(x_T)$ with respect to
$x_T$, at fixed values of $p_T$:
\begin{equation}
{\widetilde \Sigma}_{a_1a_2\to a_3, N}(p_T^2, \mu_F, \mu_f)
\equiv \int_0^1 dx_T\; x_T^{N-1}
\;{\widetilde \Sigma}_{a_1a_2\to a_3}(x_T;p_T^2, \mu_F, \mu_f) \;\;.
\end{equation}
The $N$ moments can equivalently be defined 
\cite{Catani:1998tm, deFlorian:2005yj}
with respect to $x_T^2$ (rather than $x_T$), and the two definitions are directly
related by $N \leftrightarrow 2N$.

To relate the near-threshold behaviour of the cross sections in 
Eqs.~(\ref{Sigmadef}) and (\ref{Sigmadefeta}), the main observation 
\cite{Catani:1998tm, Sterman:2000pt, Bonciani:2003nt, deFlorian:2005yj} 
is that the limit $x_T \to 1$ kinematically forces the parton rapidity
to $\eta \to 0$ (see Eq.~(\ref{xtvseta})). The function 
${\widetilde \Sigma}(x_T)$ in Eq.~(\ref{Sigmadefeta}) is obtained by the rapidity
integration of Eq.~(\ref{Sigmadef}), and we have
\beq
\label{resderiv}
{\widetilde \Sigma}(x_T)
= \int d\eta \;\;
|{\overline {{\cal M}_{a_1a_2a_3a_4}^{(0)}(r=e^{2\eta})}}|^2
\;\Theta(1 - x_T \cosh \eta) \;
\Sigma(\xw=x_T\cosh \eta, r=e^{2\eta})
\;\;,
\eeq
where we have omitted the subscript ${a_1a_2\to a_3}$ and the common dependence
on
the variables $p_T^2, \mu_F, \mu_f$, since they do not depend on the integration
variables $\eta$ and (after Mellin transformation) $x_T$. In the limit
$x_T \to 1$, considering the right-hand side of Eq.~(\ref{resderiv}),
the smooth (non-singular) dependence of 
$\Sigma(\xw, r)$ on $r=e^{2\eta}$ can be approximated by setting $\eta=0$ and,
thus,  $r=1$ (in $N$-moment space, this approximation amounts to neglecting
high-order perturbative terms of ${\cal O}(1/N)$). Then $\Sigma$ only depends on 
$\xw=x_T/x$ and this dependence enters into Eq.~(\ref{resderiv}) with the typical
convolution structure with respect to the variable $x=1/\cosh \eta$. This
convolution structure is exactly diagonalized by considering the $N$ moments,
and we directly obtain the all-order resummation formula for 
${\widetilde \Sigma}_{a_1a_2\to a_3, N}$:
\beq
\label{etaintres}
{\widetilde \Sigma}_{a_1a_2\to a_3, N}(p_T^2, \mu_F, \mu_f) =
{\widetilde \Sigma}_{a_1a_2\to a_3, N}^{(0)} 
\left[
\Sigma_{a_1a_2\to a_3 a_4,N}^{\,{\rm res}}(r=1;p_T^2,\mu_F,\mu_f) 
+ {\cal O}\left(\f{1}{N}\right)
\right] \;\;,
\eeq
where ${\widetilde \Sigma}_{a_1a_2\to a_3, N}^{(0)}$ is the  $N$ moment of the
Born-level term in Eq.~(\ref{bornetaint}).

Setting $r=1$ in our resummation formula (\ref{resu}) for 
$\Sigma_{N}^{\,{\rm res}}$, we have checked that the result in 
Eq.~(\ref{etaintres}) is consistent with the NLL resummed results for 
${\widetilde \Sigma}_{N}$ that are derived in Ref.~\cite{deFlorian:2005yj}.
In particular, since the first-order anomalous dimension ${\bom
\Gamma}^{(1)}(r=1)$ at $r=1$ involves the sole colour correlation operator
${\bom T}^2_s$ (see Eq.~(\ref{gamma190})), it can be easily diagonalized in
colour space (the eigenvectors $|I\rangle$ are the colour states of the
irreducible representations of $SU(N_c)$ that are formed by the $s$-channel
parton pair $\{a_1,a_2\}$).
The NLL resummation formula of Ref.~\cite{deFlorian:2005yj} is indeed directly
presented
in its explicitly diagonalized form (i.e., in the same form as in 
Eq.~(\ref{exponfin})). We note that the results in Ref.~\cite{deFlorian:2005yj}
neglect the colour interference between ${\bom \Delta}^{\rm (int)}_N$
and $|{\cal M}_H^{(1)}\rangle$ and (in practice) use the approximation
in Eq.~(\ref{nointapp}) (thus, the first-order contribution of 
$|{\cal M}_H|^2$ was directly extracted from the NLO results of 
Ref.~\cite{Aversa:1988vb}): this 
is 
a consistent approximation up to NLL accuracy.
We also note a difference between Eq.~(\ref{etaintres}) and the resummed
expressions of Ref.~\cite{deFlorian:2005yj}. The resummed factor 
$\Sigma_{N}^{\,{\rm res}}$ in Eq.~(\ref{etaintres}) depends on the transverse
momentum $p_T$ of the
triggered parton $a_3$, whereas the expressions of Ref.~\cite{deFlorian:2005yj}
depends on $P_{3T}=x_3p_T$ of the observed hadron $h_3$
($x_3$ is the momentum fraction of the fragmentation function $d_{a_3/h_3}$
in Eq.~(\ref{factorization})). This difference is of 
${\cal O}(1/N)$ close to the hadronic threshold, and it is an effect beyond the
LL level close to the partonic threshold. The dependence of 
$\Sigma_{N}^{\,{\rm res}}$ on $p_T$ is due to QCD scaling violation, and it 
occurs through logarithmic terms 
$\ln (p_T^2/\mu^2) = \ln (P_{3T}^2/x_3^2\mu^2)$ with $\mu=\mu_R,\mu_F,\mu_f$
(see, e.g., the NLO result in Eq.~(\ref{C1ris})).
These logarithmic terms appear in the resummation formula as coefficients of 
$\ln N$ contributions at the NLL level (and higher-order levels), 
and their effect is
comparable to the effect produced by variations of the scales 
$\mu=\mu_R,\mu_F$ and $\mu_f$.

The soft-gluon resummation formula (\ref{etaintres}) is valid to all orders.
It directly expresses the soft-gluon resummation for the rapidity-integrated
cross section in terms of the corresponding rapidity-dependent radiative factor
$\Sigma_{N}^{\,{\rm res}}$ evaluated at $r=e^{2\eta}=1$. Our result for 
$|{\cal M}_H^{(1)}\rangle$ is a necessary information to explicitly extend the 
rapidity-integrated resummation formula beyond the NLL accuracy.

\section{Summary}
\label{sec:summa}

In this paper we have considered the single-particle inclusive cross section 
at large transverse momentum in hadronic collisions.
We have studied the corresponding partonic cross section in the threshold limit
in which the final-state system that recoils against the triggered parton is
constrained to have a small invariant mass.
In this case the accompanying QCD radiation is forced to be soft 
and/or collinear and the cancellation between virtual and real infrared 
singular contributions is unbalanced, leading to large logarithmic terms 
in the coefficients of the perturbative expansion. 
Using soft and collinear approximations of the relevant five-parton matrix 
elements, we have computed the general structure of these 
logarithmically-enhanced terms in colour space at NLO.
The result of this NLO computation agrees with previous (colour summed) results
in the literature,
and it is presented here in a compact and process-independent form.
This form is factorized in colour space and this allows us to explicitly 
disentangle colour interference effects.
We have then discussed the structure of the logarithmically-enhanced terms 
beyond NLO, and
we have presented the resummation formula (see Eq.~(\ref{resu})) that
controls these contributions to the $p_T$-dependent cross section
at fixed rapidity.
The formula, which is valid at arbitrary logarithmic accuracy, is written 
in terms of
process-independent radiative factors and of a colour-space radiative 
factor ${\bom \Delta}^{\rm (int)}$ that takes into account 
soft-gluon radiation at large angles. The radiative factor 
${\bom \Delta}^{\rm (int)}$ exponentiates a colour-space anomalous dimension
${\bom \Gamma}$, whose first-order term ${\bom \Gamma}^{(1)}$ is presented in
explicit and simple form (see Eq.~(\ref{gamma1})). All the radiative factors
are explicitly given up to NLL accuracy.
Our 
process-independent
NLO result (see Eq.~(\ref{C1ris})) agrees with the expansion of the resummation
formula at the same perturbative order, and it allows us to extract the 
explicit form of the (IR finite)
hard-virtual amplitude $|{\cal M}_H^{(1)}\rangle$ at 
relative ${\cal O}(\as)$
(see Eqs.~(\ref{MH1}) and (\ref{IH1})).
This ingredient permits full control of the colour interferences in 
the evaluation of the resummation factor
$\langle {\cal M}_H|{\bom \Delta}^{\rm (int)} |{\cal M}_H\rangle$ and,
therefore, it paves the way to the explicit extension of the 
resummation formula to NNLL accuracy.
These resummation results are valid for both spin-unpolarized and 
spin-polarized hard scattering.

In the paper we have limited ourselves to considering the single-inclusive 
hadronic cross section.
The methods applied here can be used to study other important processes 
that are driven by four-parton hard scattering,
such as jet and heavy-quark production.

\noindent {\bf Acknowledgements.}
We would like to thank Daniel de Florian and Werner Vogelsang for comments on the manuscript.
This research was supported in part by the Swiss National Science Foundation (SNF) under contract 200021-144352 and by 
the Research Executive Agency (REA) of the European Union under the Grant Agreement number PITN-GA-2010-264564 ({\it LHCPhenoNet}).

\end{document}